\documentclass[12pt,a4paper]{article}

\usepackage[noblocks]{authblk}
\usepackage{a4wide,times}
\title{Bond Graph Modelling of  Chemiosmotic Biomolecular
  Energy Transduction}
\author{Peter J. Gawthrop\footnote{Corresponding author. \textbf{peter.gawthrop@unimelb.edu.au}}}
\affil{
  Systems Biology Laboratory,
  Melbourne School of Engineering,
  University of Melbourne,
  Victoria 3010, Australia.
  \authorcr
  Department of Electrical and Electronic Engineering, 
  Melbourne School of Engineering,
  University of Melbourne,
  Victoria 3010, Australia.
   }

\usepackage[authoryear,round,sort]{natbib}
\newcommand{\elec}{\text{e}^{-}}
\newcommand{\Vo}{\text{V}}

\newcommand{\FV}{\text{\(\mathbb{V}\)}}
\newcommand{\FA}{\text{\(\mathbb{A}\)}}

\newcommand{\ADP}{\text{ADP}^{3-}}
\newcommand{\ATP}{\text{ATP}^{4-}}

\newcommand{\HPO}{\text{HPO}_4^{2-}}

\newcommand{\POOO}{\text{PO}_3^{ -}}

\newcommand{\Feox}{\text{Fe}^{3+}}
\newcommand{\Fered}{\text{Fe}^{2+}}
\newcommand{\Fe}{\text{Fe}}

\newcommand{\HHO}{\text{H}_2\text{O}}

\newcommand{\HHOO}{\text{H}_2\text{O}_2}
\newcommand{\HH}{\text{H}^{+}}
\newcommand{\Hi}{\text{H}_i^{+}}
\newcommand{\Hx}{\text{H}_x^{+}}

\newcommand{\NADH}{\text{NADH}}
\newcommand{\NAD}{\text{NAD}^{+}}
\newcommand{\OO}{\text{O}_2}
\newcommand{\sOO}{\text{O}^{\bullet -}_2}
\newcommand{\PP}{\text{P}}
\newcommand{\QH}{\text{QH}_2}

\newcommand{\Q}{\text{Q}}
\newcommand{\ee}{\text{e}^{-}}
\usepackage{amsmath,amssymb,amscd,amstext,mathtools,extarrows}
\newcommand{\xt}{\tilde{x}}
\newcommand{\qt}{\tilde{q}}
\newcommand{\phit}{\tilde{\phi}}
\newcommand{\Phit}{\tilde{\Phi}}
\newcommand{\Phib}{{\bar{\Phi}}}
\newcommand{\Faraday}{\mathbf{F}}
\newcommand{\pmf}{\Delta p}
\newcommand{\emf}{\Delta \psi}

\newcommand{\Std}{\ominus}
\newcommand{\std}{\oslash}

\newcommand{\lb}{\left (}
\newcommand{\rb}{\right )}
\newcommand{\reacul}[2]{
  {\; \xrightleftharpoons[#2]{#1} \;}
}

\newcommand{\reacu}[1]{
  \reacul{#1}{}
}

\newcommand{\reac}{
  \reacu{}
}

\usepackage{siunitx}
\newcommand{\BG}[1]{\text{\sffamily\textbf{#1}}}

\newcommand{\C}{\BG{C }}

\newcommand{\R}{\BG{R }}

\renewcommand{\SS}{\BG{SS }}

\newcommand{\TF}{\BG{TF }}

\renewcommand{\Re}{\BG{Re }}

\newcommand{\BGL}[2]{$\BG{#1}$:$\mathbf{#2}$} 

\newcommand{\BC}[1]{\BGL{C}{#1}}

\newcommand{\BSS}[1]{\BGL{SS}{#1}}

\newcommand{\BTF}[1]{\BGL{TF}{#1}}

\newcommand{\BRe}[1]{\BGL{Re}{#1}}

\usepackage{listings}

\usepackage{graphicx,subfigure,fancybox,color}

\newcommand{\Fig}[2]{
  \includegraphics[width=#2\linewidth]{#1.pdf}
   \label{subfig:#1}
}

\newcommand{\SubFig}[3]{
  \subfigure[#2]{
    \includegraphics[width=#3\linewidth]{#1.pdf}
    \label{subfig:#1}
  }
}

\usepackage{doi}
\usepackage{url}

\begin{document}
\maketitle

\begin{abstract}
  Engineering systems modelling and analysis based on the bond graph
  approach has been applied to biomolecular systems.  In this
  context, the notion of a \emph{Faraday-equivalent} chemical
  potential is introduced which allows chemical potential to be
  expressed in an analogous manner to electrical volts thus allowing
  engineering intuition to be applied to biomolecular systems.
  Redox reactions, and their representation by half-reactions, are key
  components of biological systems which involve both electrical and
  chemical domains. A bond graph interpretation of redox
  reactions is given which combines bond graphs with the
  Faraday-equivalent chemical potential.
  This approach is particularly relevant when the biomolecular system
  implements chemoelectrical transduction -- for example chemiosmosis
  within the key metabolic pathway of mitochondria: oxidative
  phosphorylation.
   
  An alternative way of implementing computational modularity using
  bond graphs is introduced and used to give a physically based model
  of the mitochondrial electron transport chain (ETC).
  To illustrate the overall approach, this model is analysed using the
  Faraday-equivalent chemical potential approach and engineering
  intuition is used to guide \emph{affinity equalisation}: a energy
  based analysis of the mitochondrial electron transport chain.
\end{abstract}

\section{Introduction}
\label{sec:introduction}
Like engineering systems, living systems are subject to the laws of
physics in general and the laws of thermodynamics in particular.
This fact gives the opportunity of applying engineering approaches to
the modelling, analysis and understanding of living systems.
The bond graph method of \citet{Pay61} is one such well-established
engineering approach~\citep{Cel91,GawSmi96,GawBev07,Bor10,KarMarRos12}
which has been extended to include biomolecular
systems~\citep{OstPerKat71,OstPerKat73}.
To quote from \citet{Pay93a}:
\begin{quotation}
  \emph{Katchalsky's breakthroughs in extending bond graphs to
    biochemistry are very much on my own mind. I remain convinced that
    BG models will play an increasingly important role in the upcoming
    century, applied to chemistry, electrochemistry and biochemistry,
    fields whose practical consequences will have a signiﬁcance
    comparable to that of electronics in this century. This will occur
    both in device form, say as chemfets, biochips, etc, as well as in
    the basic sciences of biology, genetics, etc.}
\end{quotation}
With this quotation in mind, this paper builds on the pioneering work
of Katchalsky's group
\citep{OstPerKat71,OstPerKat73}, together with more recent
investigations  \citep{GawCra14,GawCurCra15,GawCra16,Gaw17} to give
an engineering-inspired modelling approach to biomolecular systems
which seamlessly combines biochemical reactions, electrons and protons
using the concept of the \emph{ Faraday-equivalent} chemical potential.
In particular, this paper shows that combining electrical units for
chemical potential with bond graph models of biomolecular systems not
only provides a systematic methods for model development and analysis
of biomolecular systems but also provides a bridge allowing
application of electrical engineering methodology to biomolecular
networks.

Redox reactions provide the energy required to sustain
life~\citep{AtkPau11,SouThiLan13} and the notion of the \emph{redox
  potential} is useful in describing energetic properties. This paper
shows that both redox reactions and redox potential can be clearly and
explicitly described using the bond graph approach and the use of the
Faraday-equivalent chemical potential.

Mitochondria make use of redox reactions to provide the power driving
many living systems. Mathematical modelling of the key components of
mitochondria is thus an important challenge to systems biology.
As discovered by \citet{Mit61,Mit76,Mit93,Mit11}, the key feature of
mitochondria is the \emph{chemiosmotic} energy transduction whereby a
chain of redox reactions pumps protons across the mitochondrial inner
membrane to generate the \emph{proton-motive force} (PMF). This PMF is
then used to power the synthesis of ATP -- the universal fuel of
living systems.
Because mitochondria transduce energy, an energy-based modelling
method \citep{Hil89,QiaBea05,WuYanVin07,BeaQia10} is desirable.

Modular bond graphs provide a way of decomposing complex biomolecular
systems into manageable parts~\citep{GawCurCra15,GawCra16}. This paper
combines the modularity concepts of \citet{NeaCarTho16} with the bond
graph approach to give a more flexible approach to modularity.
This paper suggests that such a modular bond graph approach, combined with
electrical units, provides a flexible and powerful energy-inspired
modelling method which brings engineering expertise to the analysis of
biomolecular systems in general and chemiosmotic energy transduction
in mitochondria in particular.

An alternative approach would use electrical networks to model
chemical systems
\citep{OstPer74,ZupPasJur04,CarSanSot14}. Indeed,
\citet{OstPer74} show the precise connection between the two
approaches. However, the resultant circuit diagrams can be unwieldy
and the representation of stoichiometry is cumbersome. Therefore, in
the author's opinion, the more general bond graph approach is
superior. Nevertheless, the equivalence discussed by \citet{OstPer74}
should, in principle, allow circuit-theoretical
approaches~\citep{AndVon06} to be incorporated.

\S~\ref{sec:farad-equiv-potent} introduces the Faraday-equivalent
chemical potential and this is used in \S~\ref{sec:redox-reactions} to
provide bond graph models of redox reactions which seamlessly combine
the chemical and electrical domains and provide a bond graph
interpretation of redox potential.
\S~\ref{sec:modularity} considers an approach to computational
modularity in the context of bond graphs which is then used, together
with the redox reaction models of \S~\ref{sec:redox-reactions}, in
\S~\ref{sec:electr-transp-chain} to give a modular bond graph model of
the mitochondrial electron transport chain (ETC). \S~\ref{sec:equalise} uses
this bond graph model to analyse how the intermediate electron
transporters coenzyme Q and cytochrome c equalise the
Faraday-equivalent potentials along the mitochondrial electron
transport chain.
\S~\ref{sec:ATP} describes how the bond graph representation of redox
reactions can be generalised to include ATP hydrolysis and synthesis
and how this can be combined with the ETC to give a modular bond graph
representation of oxidative phosphorylation.
\S~\ref{sec:conclusion} concludes the paper and suggests directions
for future research.

\section{The Faraday-equivalent potential}
\label{sec:farad-equiv-potent}
The fundamental biophysical processes of life involve the transduction
of chemical energy and electrical energy \citep{LanMar10}. For example,
the chemiosmotic theory of \citet{Mit61,Mit76,Mit93,Mit11} explains how a
mixture of chemical and electrical energy is stored in a trans-membrane
proton gradient and the theory of \citet{HodHux52} shows how the
mutual transduction of chemical and electrical energy gives rise to
action potential in nerves.

Because the chemical and electrical domains are so intertwined, the
analysis and understanding of such systems is enhanced by a common
approach to the two domains. One example of this is the \emph{proton
  motive force} PMF of chemiosmotic theory
\citep{Mit93,BerTymStr12,NicFer13,AlbJohLew15} which reexpresses the
chemical potential of protons as electrical voltage using the Faraday
constant so that it can be added to the electrical potential.  A
second example is the notion of \emph{redox
  potential}~\citep{AtkPau11,BerTymStr12,AlbJohLew15} which assigns a
voltage to reactions involving electron transfer.

A theme of this paper is that the notion of rexpressing chemical
potential as electrical potential is not just confined to
electrically-charged ions but can be generally applied to any chemical
species -- charged or not. Indeed, this can be regarded as one aspect
of the concept of physical analogies introduced by \citet{Max71} who
pointed out that analogies are central to scientific thinking and
allow mathematical results and intuition from one physical domain to
be transferred to another. The central concept here is that conservation of
\emph{energy} holds across different physical domains.

\subsection{Variables \& Units}
In the context of electrochemical systems, there are two ways of
unifying the two domains: reexpress chemical potential as electrical
potential \citep{BosFreEva03} (as in the proton-motive force concept
\citet{Mit93,Mit11}) or reexpress electrical potential as chemical
potential \citep{GawSeiKam15X}.  Those with a physics or
engineering background would be more familiar with electrical units
and would therefore prefer the former choice. However there is a more
general reason for choosing the electrical domain: it is better
endowed with dedicated units.

Chemical potential is expressed as the compound unit of Joules per mole
(\si{J.mol^{-1}}) but does not have a dedicated unit%
\footnote{
\citet{JobHer06} suggest the Gibbs (\si{G}) as a the unit of  chemical
potential, but this is not widely used.
}. 
In contrast, electrical potential has its own unit, the Volt
(\si{V}). Although it would be possible to ignore this unit and use
the equivalent compound unit of Joules per Coulomb (\si{J.C^{-1}}), this
would obscure the basic simplicity of electrical theory.
Moreover, chemical flow can be expressed in compound units as moles
per second
(\si{mol.s^{-1}}) but does not have a dedicated unit; in contrast,
electrical flow has its own unit, the Amp (\si{A}). Again, it would be
possible to be perverse and ignore this unit and use the equivalent
compound unit of Coulombs per second (\si{C.s^{-1}}).

The conversion factor relating the electrical and chemical domains is
\emph{Faraday's constant} $\Faraday\approx\SI{96485}{C.mol^{-1}}$. As
discussed by \citet{Kar90} and \citet{GawSeiKam15X}, this conversion
can be represented by the bond graph \TF component
which enforces energy conservation. 
Like all physical quantities, Faraday's constant $\Faraday$ is
composed of a real number (the \emph{measure}) and a \emph{unit}
\citep{Wat96}. In particular:
\begin{align}
  \Faraday &= F \times  U\\
  \text{where } F  &\approx 96485\\
\text{and } U &=  \SI{1}{C.mol^{-1}}
\end{align}
In bond graph  terms, the single bond graph  \TF component
representing $\Faraday$ has been split into two \TF components: one
representing the the purely numerical conversion $F$ and one
representing the purely dimensional conversion $U$.

Hence it is possible to define two new \emph{derived} units, the
\emph{Faraday-equivalent voltage} $\si{\FV} = F\si{J.mol^{-1}}$ and the
\emph{Faraday-equivalent current} $\si{\FA} = \frac{1}{F}\si{mol.s^{-1}}$.
Using these units, the \emph{Faraday-equivalent potential} $\phi$,
\emph{Faraday-equivalent affinity} $\Phi$ and the
\emph{Faraday-equivalent flow} $f$ are defined in terms of chemical
potential $\mu$ and molar flow $v$ as:
\begin{xalignat}{2}
  &\text{Faraday-equivalent chemical potential}& \phi &= \frac{\mu}{F}\si{~\FV}\label{eq:phi}\\
  &\text{Faraday-equivalent reaction affinity}& \Phi &= \frac{A}{F}\si{~\FV} \label{eq:Phi}\\
  &\text{Faraday-equivalent flow}& f &= F v \si{~\FA}\label{eq:f}
\end{xalignat}
For example, consider NAD at standard conditions which has a chemical
potential at standard conditions $\mu_{NAD}^\Std = \SI{18100}{J.mol^{-1}}$; the corresponding
Faraday-equivalent potential is $\phi_{NAD}^\Std =
\SI{188}{\milli\FV}$. Similarly, a molar flow of $v=\SI{1}{\micro mol.s^{-1}}$ has a
Faraday-equivalent flow of about $f=\SI{97}{\milli\FA}$. 
Faraday-equivalent chemical potentials for some other species are
given in Table \ref{tab:FEP}.

\subsection{The Bond Graph  \C component}
The \C component is the bond graph  abstraction of an electrical
capacitor. In the chemical context, it represents a chemical species
with chemical potential replacing voltage and molar flow replacing
current \citep{OstPerKat71,OstPerKat73}.
In particular, following \citet{GawCra14}, the bond
graph \C component for biomolecular systems accumulates a chemical
species A as the number of moles $x_A$ and generates the corresponding
chemical potential $\mu_A$ in terms of the molar flow $x_A$:
\begin{align}
  x_A(t) &= \int_0^t v_A(t^\prime) dt^\prime  + x_A(0)\label{eq:x_A}\\
  \mu_A &= \mu_A^\std + RT \ln \frac{x_A}{x_A^\std} \label{eq:mu_A}
\end{align}
where $\mu_A^\std $ is the chemical potential of $x_A$ when $x_A=x_A^\std$.
Equation \eqref{eq:mu_A} may be rewritten in two ways:
\begin{align}
  \mu_A &= RT \ln K_A x_A \\
  \text{where } K_A &= \frac{e^\frac{\mu_A^\std}{RT}}{x_A^\std} \label{eq:mu_A_K}\\
  \text{and } 
  \mu_A &= \mu_A^\std + RT \ln \lb 1 + \frac{\xt_A}{x_A^\std} \rb \\
\text{where }  \xt_A &= x_A - x_A^\std \label{eq:mu_A_tilde}
\end{align}
Equation \eqref{eq:mu_A_K} is equivalent to that used previously
\citep{GawCra14,Gaw17} and equation \eqref{eq:mu_A_tilde} is
convenient when $\xt_A$ is small and so:
\begin{xalignat}{2}
  \mu_A &\approx \mu_A^\std + RT\frac{\xt_A}{x_A^\std} &
  \text{when } \frac{\xt_A}{x_A^\std} &\ll 1
\end{xalignat}

Using equations \eqref{eq:phi} and \eqref{eq:f}, equations \eqref{eq:x_A} and
\eqref{eq:mu_A} can be rewritten in Faraday-equivalent form as:
\begin{align}
   q_A(t) &= \int_0^t q_A(t^\prime) dt^\prime  + q_A(0) \label{eq:q_A}\\
  \phi_A &= \phi_A^\std + \phit_A\label{eq:phi_A}\\
\text{where }  
\phi_A^\std &= \frac{\mu_A^\std}{F}\\
\phit_A &= V_N \ln \frac{q_A}{q_A^\std} 
           = V_N \ln \lb 1 + \frac{\qt_A}{q_A^\std} \rb\label{eq:tphi_A}\\
V_N &= \frac{RT}{F} \approx \SI{26}{\milli\FV}\label{eq:V_N}\\
q_A &= F x_A\\
q_A^\std &= F x_A^\std\\
\text{and } \qt_A &= q_A - q_A^\std
\end{align}

\begin{table}[htbp]
  \centering
  \begin{tabular}{|l||l|l|}
\hline
Species &$\mu^\Std\si{kJ.mol^{-1}}$ & $\phi^\Std\si{\FV}$\\
\hline
\hline
$\OO$ & 16.4 & 0.169974\\
$\HH$ & 0    & 0  \\
$\HHO$ & -235.74 & -2.44327\\
$\NADH$ & 39.31 & 0.407419 \\
$\NAD$ & 18.1 & 0.187593\\
$\Q$ & 65.17 & 0.675439\\
$\QH$ & -23.3 & -0.241487\\
$\Feox$ & -6.52 & -0.067575\\
$\Fered$ & -27.41 & -0.284085\\
$\ATP$ & -2771 & -28.7194\\
$\ADP$ & -1903.96 & -19.7332\\
$\HPO$ & -1098.27 & -11.3828\\
$\HH_x$ (pH 7.78) & -44.408& -0.46026  \\
\hline
  \end{tabular}
  \caption[Faraday-equivalent Potentials]
  {Chemical and Faraday-equivalent
    Potentials. }
\label{tab:FEP}
\end{table}
Tables of chemical potentials at standard conditions are available
\citep{AtkPau11}.  Table \ref{tab:FEP} lists some of these with their
Faraday-equivalent potentials where the values for $\mu^\Std$ are
taken from \citet{WuYanVin07}.
Given the chemical potential of substance A at
standard conditions $\mu^\Std_A$, and the corresponding
Faraday-equivalent potential $\phi^\Std_A=F\mu^\Std_A$, the
Faraday-equivalent potential $\phi^\std_A$ at any other operating
point can be computed from
Equation \eqref{eq:phi_A} as
\begin{align}
  \phi^\std_A &= \phi^\Std_A + V_N \ln \rho_A\\
\text{where } \rho_A &= \frac{q_A^\std}{q_A^\Std} = \frac{c_A^\std}{c_A^\Std}
\end{align}
and $c_A^\std$ and $c_A^\Std$ are the concentrations at the relevant
conditions. For example, the following are required in \S~\ref{sec:equalise}.
\begin{xalignat}{2}
  &\text{$\HH$ at pH 7} & \phi^\std &= 0 + V_N \ln 10^{-7}  =\SI{-414}{\milli\FV}\\
  &\text{$\Hi$ at pH 6.88} & \phi^\std &= 0 + V_N \ln 10^{-6.88}  =\SI{-407}{\milli\FV}\\
  &\text{$\Hx$ at pH 7.78} & \phi^\std &= 0 + V_N \ln 10^{-7.78}  =\SI{-460}{\milli\FV}\label{eq:phi_Hx}\\
  &\text{$\OO$ at \SI{200}{\micro M} } & \phi^\std  &= 170 + V_N \ln
  \text{2e-4}  =\SI{-49}{\milli\FV} 
\end{xalignat}
The pH values for $\HH_x$ and $\HH_i$ are taken from
\citet{PorGheZan05}.
  
\subsection{Chemostats}
\label{sec:chemostats}
%
As discussed by \citet{GawCra16}, the notion of a \emph{chemostat}
\citep{PolEsp14} is useful in creating an open system from a closed
system; a similar approach is used by \citet{QiaBea05} who use the
phrase ``concentration clamping''.
The chemostat has four interpretations:
\begin{enumerate}
\item one or more species is fixed to give a constant concentration
  \citep{GawCurCra15}; this implies that an appropriate external
  flow is applied to balance the internal flow of the species.
\item an ideal feedback controller is applied to species to be fixed
  with setpoint as the fixed concentration and control signal an
  external flow.
\item as a \C component with a fixed state and
\item as an ideal source of Faraday-equivalent potential: $\phi~=~\phi^\std$.
\end{enumerate}

In this paper, a further interpretation is added. In
\S~\ref{sec:modularity}, a chemostat is interpreted as an external
\emph{port} of a module which allows connection to other modules.

\subsection{The Bond Graph  \Re component}
The \R component is the bond graph abstraction of an electrical
resistor. In the chemical context, a two-port \R component represents
a chemical reaction with chemical affinity (net chemical potential)
replacing voltage and molar flow replacing current
\citep{OstPerKat71,OstPerKat73}. As it is so fundamental, this two
port \R component is given a special symbol: \Re~\citep{GawCra14}. In
particular, the \Re component determines a reaction flow $v_1$ in
terms of forward and reverse affinities $A^f_1$ and $A^r_1$ as the
\emph{Marcelin -- de Donder} formula~\citep{Rys58}:
\begin{align}
  v_1 &= \kappa_1 \lb \exp \frac{A^f_1}{RT} - \exp \frac{A^r_1}{RT} \rb\label{eq:v_1}
\end{align}
in the special case of mass-action kinetics, $\kappa$ is a
constant. Otherwise $\kappa$ is a function of the forward and reverse
affinities $A^f_1$ and $A^r_1$.  Using equations \eqref{eq:Phi} and
\eqref{eq:f}, equation \eqref{eq:v_1} can be rewritten in
Faraday-equivalent form as:
\begin{align}
  f_1 &= \frac{V_N}{r_1} \lb \exp \frac{\Phi^f_1}{V_N} - \exp
        \frac{\Phi^r_1}{V_N} \rb\label{eq:f_1}\\
\text{where } r_1 &= \frac{V_N}{F\kappa_1}
\end{align}
$V_N$ is given by Equation \eqref{eq:V_N} and the resistance $r_1$ has
units of ohms ($\si{\ohm}$).
Alternatively:
\begin{align}
  f_1 &= \frac{V_N}{r_1} 2 \exp \frac{\bar{\Phi}_1}{V_N} \sinh \frac{\frac{\Phi_1}{V_N}}{2}\label{eq:f_1_alt}\\
  \text{where } \bar{\Phi}_1 &= \frac{\Phi^f_1+\Phi^2_1}{2}\\
  \text{and } \Phi &= \frac{\Phi^f_1-\Phi^r_1}{2}
\end{align}
When the normalised reaction affinity $\frac{\Phi_1}{V_N} \ll 1$:
\begin{align}
  f_1 &\approx \lb \exp \frac{\bar{\Phi}_1}{V_N} \rb \frac{\Phi_1}{r_1}
\end{align}

\section{Redox reactions}
\label{sec:redox-reactions}
\begin{figure}[htbp]
  \centering
  \Fig{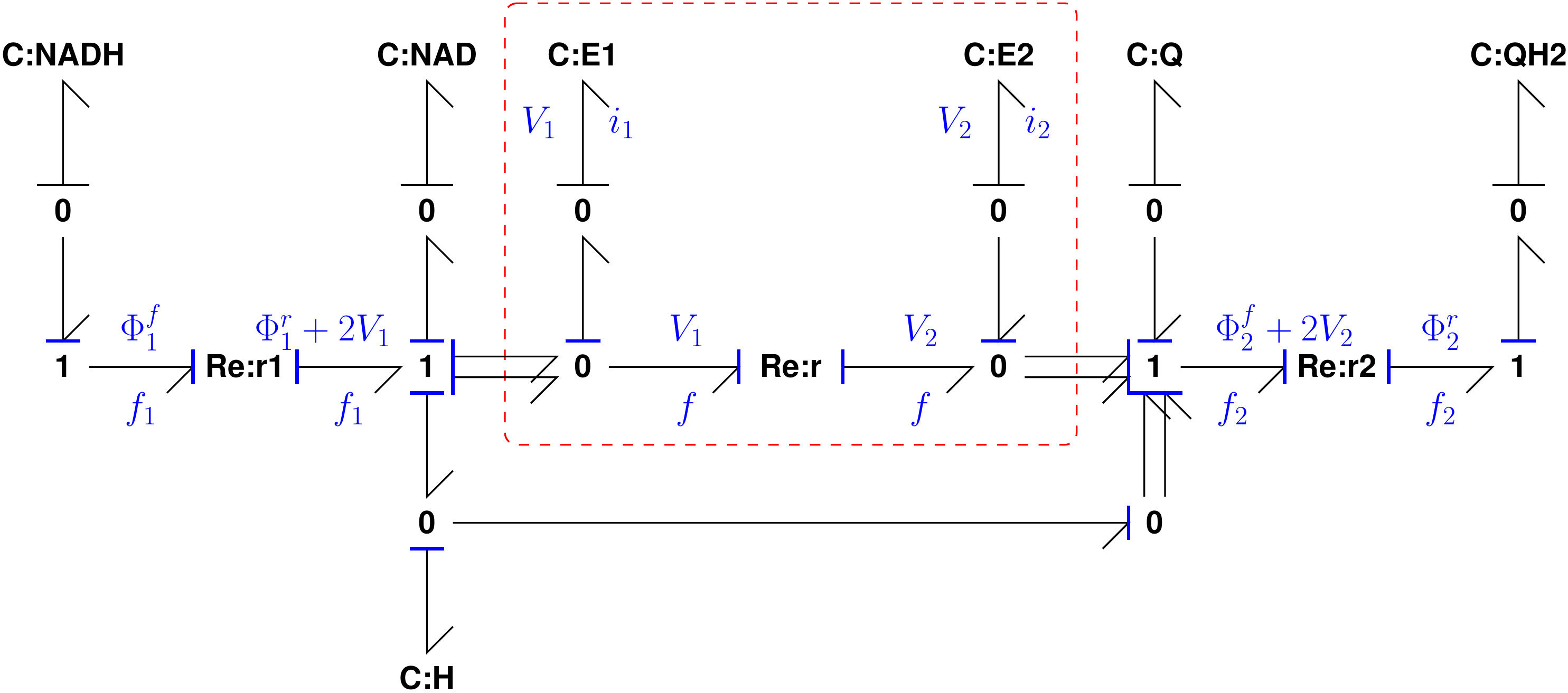}{1}
  \caption[Redox Reactions]{Redox Reactions. The redox reaction
    $\NADH + \Q + \HH \reac \NAD + \QH$ is divided into two
    half-reactions $\NADH \reacu{r1} \NAD + \HH +2\elec_1$ and
    $\Q + 2\HH + 2\elec_2 \reacu{r2} \QH$ and the electron transfer is
    represented by $\elec_1 \reacu{r} \elec_2$. The dashed box
    contains the electrical part of the system with linear components;
    the rest of the system is chemical and nonlinear.}
\label{fig:CIredox}
\end{figure}

Redox reactions \citep[Chapter 5]{AtkPau11} involve the transfer of
electrons $\elec$, and the corresponding free energy, from a donor
species to an acceptor species. This can be explicitly represented
using the concept of \emph{half-reactions} \citep[\S 5.4]{AtkPau11}.
For example in the reaction
\begin{equation}\label{eq:NADH}
  \NADH + \Q  + \HH \reac \NAD + \QH 
\end{equation}
$\NADH$ (reduced Nicotinamide Adenine Dinucleotide) donates two
$\elec$ (electrons) in forming $\NAD$ (oxidised Nicotinamide Adenine
Dinucleotide) which are accepted by $\Q$ (oxidised Ubiquinone) to form $\QH$
(reduced Ubiquinone) \citep[Panel 14.1]{AlbJohLew15}.

Reaction \eqref{eq:NADH} can be split into two half reactions as:
\begin{align}
  \NADH   &\reacu{r1} \NAD + \HH +2\elec_1 \label{eq:NADHa}\\
  \Q + 2\HH + 2\elec_2 &\reacu{r2} \QH \label{eq:NADHb}
\end{align}
where $\elec_1$ denotes electrons donated in half-reaction $a$
\eqref{eq:NADHa} and $\elec_2$ denotes electrons accepted in
half-reaction $b$ \eqref{eq:NADHb}.

Figure \ref{fig:CIredox} includes the two chemical
half-reactions \eqref{eq:NADHa} and \eqref{eq:NADHb} together with an
electrical interconnection. Thus the bond graph component \BRe{r1},
together with the components \BC{NADH}, \BC{NAD} and \BC{H} and
connecting bonds represents reaction $r1$ and the bond graph component
\BRe{r2}, together with the components \BC{Q}, \BC{QH2} and \BC{H} and
connecting bonds represents reaction $r2$.
To obtain the appropriate bond graph, it is assumed that the
electrons associated each reaction accumulate in electrical
capacitors represented by \BC{Ea} and \BC{Eb} which generate voltages
$\Vo_1$ and $\Vo_2$ respectively. The corresponding electrical currents
are:
\begin{xalignat}{2}
  i_1 &= 2f_1 - f & i_2 &= f - 2f_1
\end{xalignat}
It is further assumed that electrons can flow via the electrical
resistor \BRe{r}. This is represented by the reaction
\begin{equation}
  \elec_1 \reacu{r} \elec_2
\end{equation}
and corresponds to the current:
\begin{equation}
  f = \frac{\Vo_1 - \Vo_2}{r}
\end{equation}

The chemoelectrical redox system of Figure \ref{fig:CIredox}
spans the two physical domains (chemical and electrical) discussed in
\S~\ref{sec:farad-equiv-potent}. The standard approach to redox
potential is to view the chemical part of the system from an electrical
point of view; this is now shown to have bond graph interpretation.

In particular, consider the case where the electrical resistor is
open-circuit so that the current $f=0$. When the two separate parts of
the system are in equilibrium, the two reaction flows are zero:
$f_1=0$, $f_2=0$: this implies that the net affinity for reaction \BRe{r1}
must be exactly balanced by the voltage on the electrical capacitor
\BC{E1} and the net affinity for reaction \BRe{r2}
must be exactly balanced by the voltage on the electrical capacitor
\BC{E2}:
\begin{xalignat}{2}
  2V_1 &= \Phi_1 & 2V_2 &= \Phi_2 \label{eq:2V_1}
\end{xalignat}
Focusing on half-reaction $1$, and using Table \ref{tab:FEP} and the
potential for $\Hx$ from Equation \eqref{eq:phi_Hx}, the reaction
affinity $\Phi_1$ is given by:
\begin{align}
  \Phi_1 = \Phi^f_1 - \Phi^r_1 &= \phi^\std_{NADH} -\lb \phi^\std_{NAD} +
                                 \phi^\std_{Hx} \rb\\
                               &= 408 - \lb 188 - 460 \rb = \SI{680}{\milli\FV}
\end{align}
%
%
From Equation \eqref{eq:2V_1},
\begin{align}
  V_1 &= \frac{1}{2} \Phi_1 = \SI{340}{mV}
\end{align}
$V_1$ is the redox potential of half-reaction 1.
Similarly:
\begin{align}
  \Phi_2 &= \phi^\std_{QH2} - \phi^\std_{Q} - 2\phi^\std_{Hx} \notag\\
         &= -241 - 675 + 920 = \SI{4}{\milli\FV}\\
  V_2^\std &= \frac{1}{2} \Phi_2 
           = \SI{2}{mV}
\end{align}
Using the standard sign convention, $V_2^\std$ is minus the redox
potential of half-reaction 2. The overall redox potential is given by:
\begin{equation}
  V^\std = V_1^\std - V_2^\std = \SI{338}{mV}
\end{equation}

If current is allowed to flow through the resistor, all the energy
associated with the redox potential $V^\std$ is wastefully dissipated.
In contrast, the CI complex of the mitochondrial respiratory chain
uses the flow of electrons to pump protons across the inner
mitochondrial membrane against both a concentration and electrical
gradient: thus much of the energy associated with $V^\std$ is
transduced and stored as the mitochondrial proton-motive
force~\citep{Mit93,Mit11,NicFer13}. This is examined in \S\S
\ref{sec:modularity} \& \ref{sec:electr-transp-chain}.

\section{Computational Modularity}
\label{sec:modularity}
\begin{figure}[htbp]
  \centering
  \SubFig{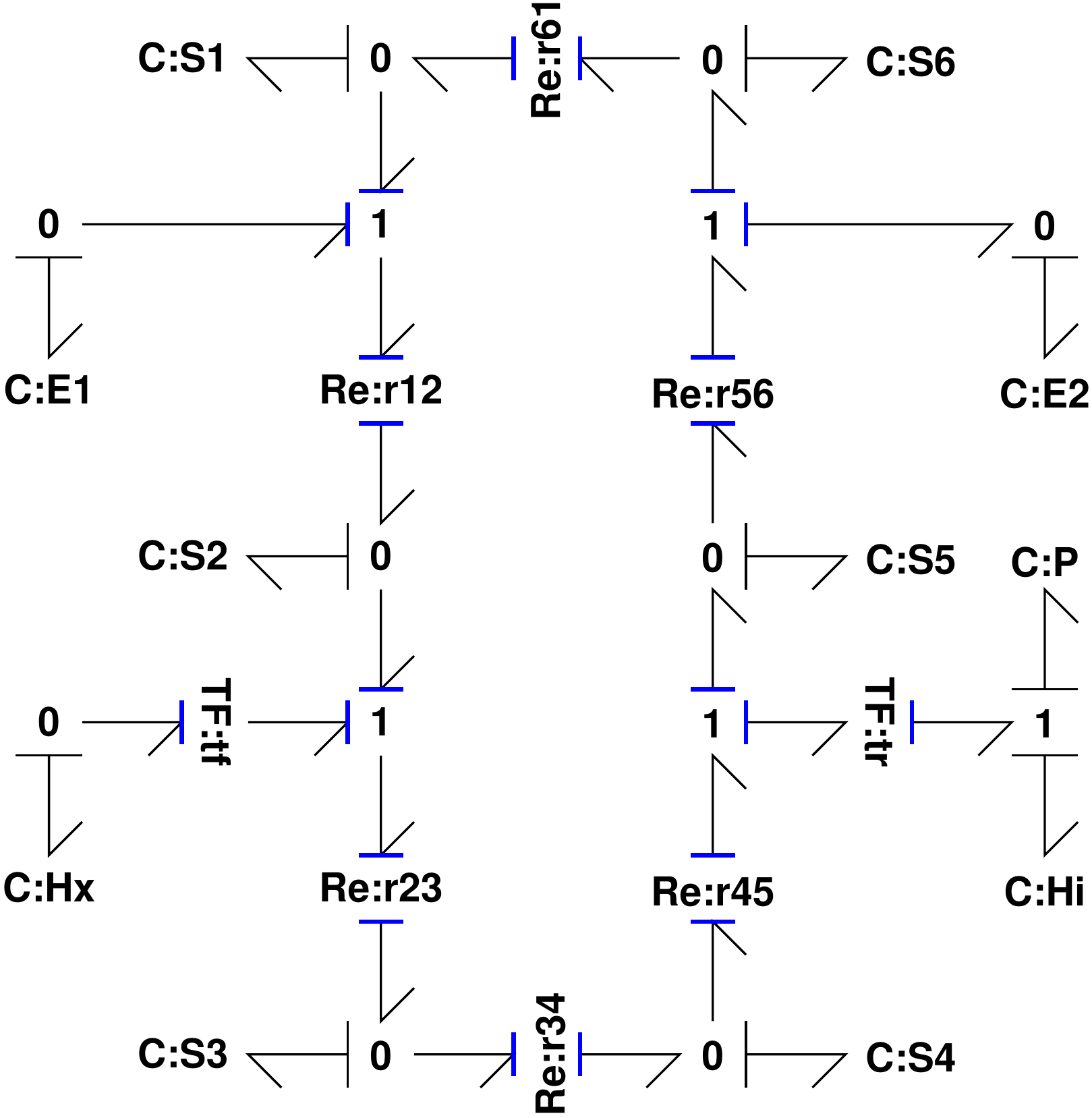}{Proton pump}{0.45}
  \SubFig{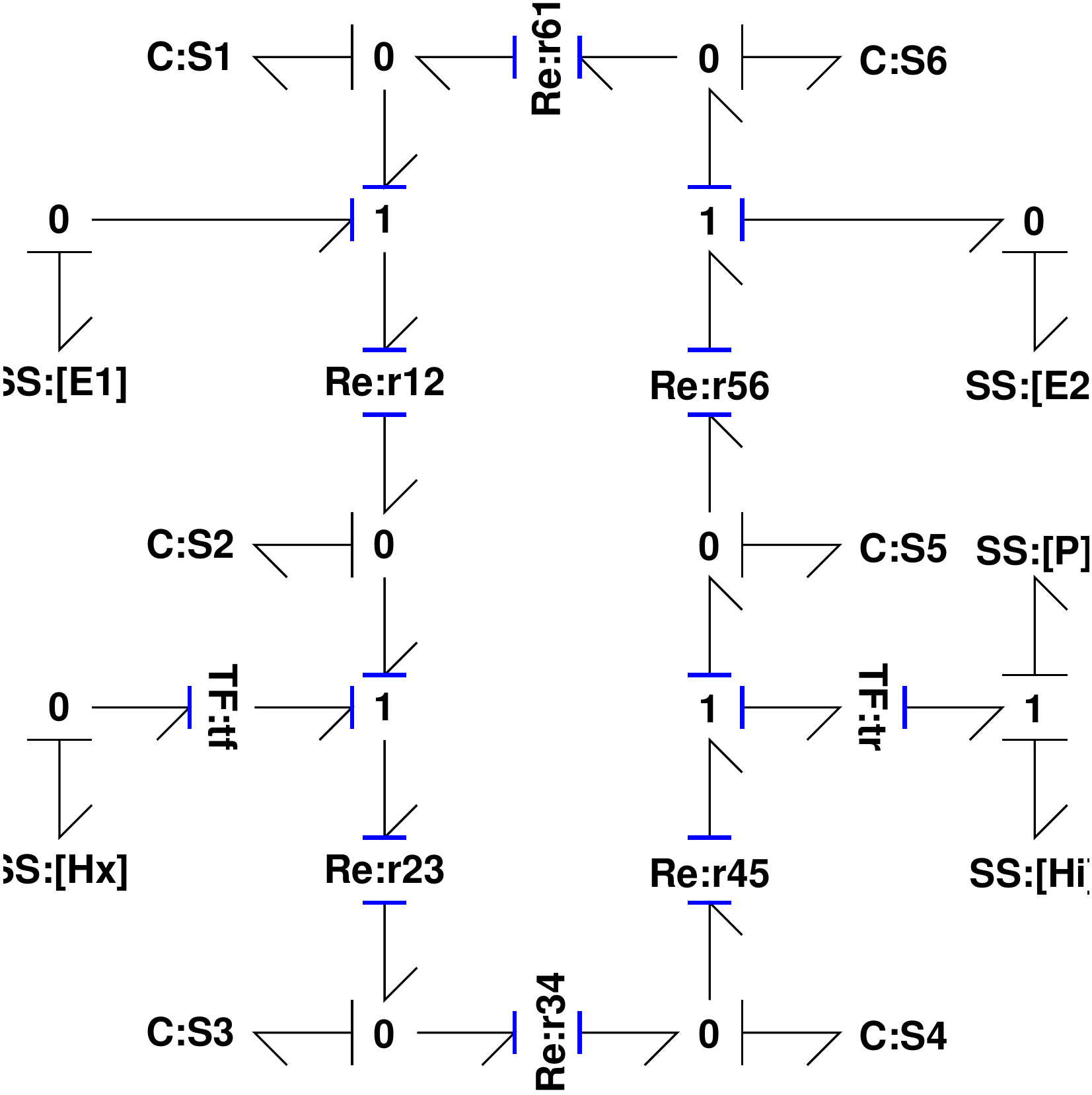}{Proton pump module}{0.45}
  \caption{Modularity and Proton Pumps. (a) A model of an
    electron-driven proton pump; the two transformers \BTF{tf} and
    \BTF{tr} have the same modulus $n_p$: the number of protons pumped
    per electron.  (b) A modular version of (a). 
%
}
\label{fig:pump}
\end{figure}
As discussed by \citet{NeaCooSmi14}, models of biological systems
should be modular and reusable. Modularity raises the issue of module
interfaces and \citet{NeaCooSmi14} distinguish between black-box,
code-level coupling using information-hiding interfaces at one extreme
and white-box, biological-level coupling at the other.

Bond graphs naturally give rise  hierarchical modular
modelling \citep{Cel91,Cel92}. One approach uses explicit ports,
represented by the bond graph  \SS~(source-sensor) component, to 
define interfaces \citep{GawBev07} and this has been used in
the biomolecular context \citep{GawCurCra15}. However, this does
have the disadvantages of the black-box approach discussed by
\citet{NeaCooSmi14}.
This paper uses an alternative approach to bond graph modularity
inspired by the approach of \citet{NeaCarTho16}. The basic idea is
simple: modules are self-contained and have no explicit ports; but any
species, as represented by a \C component has the potential to become
a port. Thus if two modules share the same species, the corresponding
\C component in each module is replaced by an \SS ~component with the
same name, and the species is explicitly represented as a \C component
on a higher level. Moreover, each module can be individually tested by
replacing the relevant \C components by chemostats.
Although not present in the current implementations, explicit
connection to ontology databases such as the Ontology of Physics for
Biology \citep{CooBooGen11} or composite ontologies
\citep{GenNeaGal11} would be required for general use.

The mitochondrial proton pumps are complex molecules
\citep{SchCha01}. Nevertheless, their key energetic features can be
modelled using simplified representations.
Figure \ref{subfig:ppn_cbg} shows a simple model of an electron-driven
proton pump based on the generic biomolecular cycle of \citet{Hil89}. The two electrical capacitors \BC{E1} and \BC{E2}
correspond to those modelling redox potential in Figure
\ref{fig:CIredox}; \BC{Hx} and \BC{Hi} correspond to the amount of
protons in the mitochondrial matrix and intermembrane space
respectively and \BC{P} to the proton electrical potential across the
membrane. \BRe{r} determines the flow of protons though the
membrane. 
In this particular case, all five \C components are used as external
connections in the modular form of the model shown in
\ref{subfig:mPPn_cbg}.
This model is reused three times (with differing values of $n_p$) in
\S~\ref{sec:electr-transp-chain} as part of the models of the CI, CIII
and CIV complexes of the mitochondrial electron transport chain (ETC).

\begin{figure}[htbp]
  \centering
  \SubFig{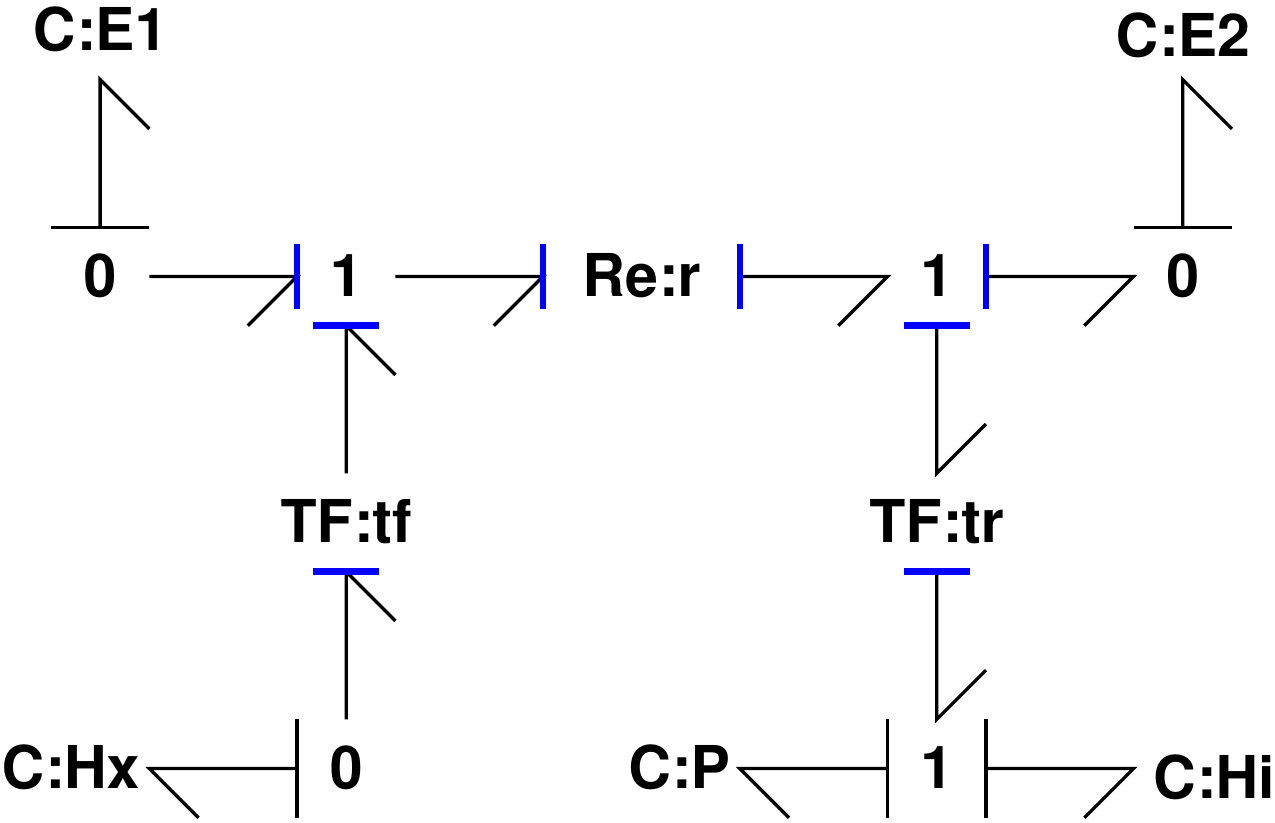}{Simplified Proton pump}{0.35}
  \SubFig{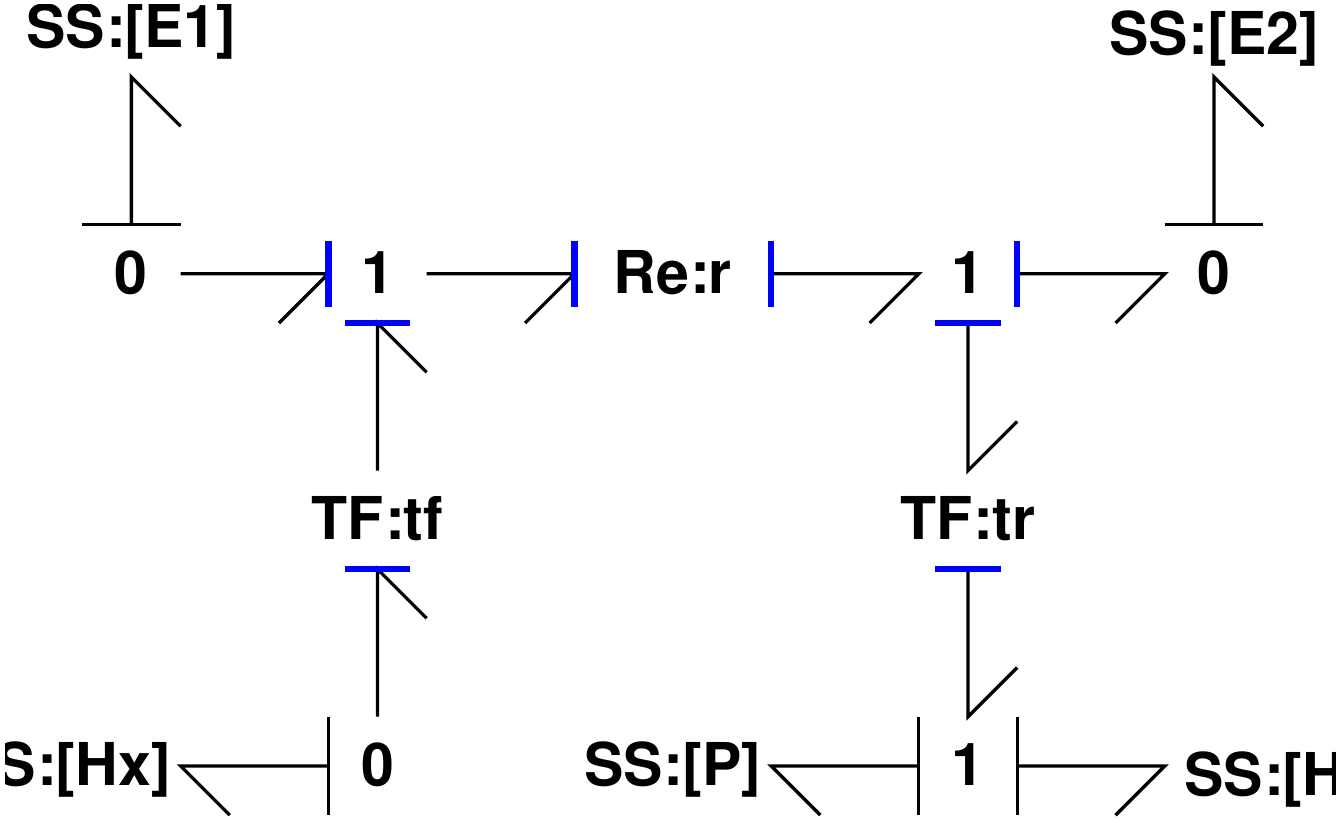}{Modular simplified Proton pump}{0.35}
  \caption{Physically-plausible simplification}\label{fig:simplified}
\end{figure}
As in any modelling endeavour, the complexity of the model should be
appropriate to its use; the important issue is to enable the modelling
approach to handle a range of levels of complexity whilst retaining a
physically correct representation.
Figure~\ref{fig:simplified} shows one possible simplification of the
module of Figure~\ref{fig:pump}.

\section{The Mitochondrial Electron Transport Chain}
\label{sec:electr-transp-chain}
\begin{figure}[htbp]
  \centering
  \Fig{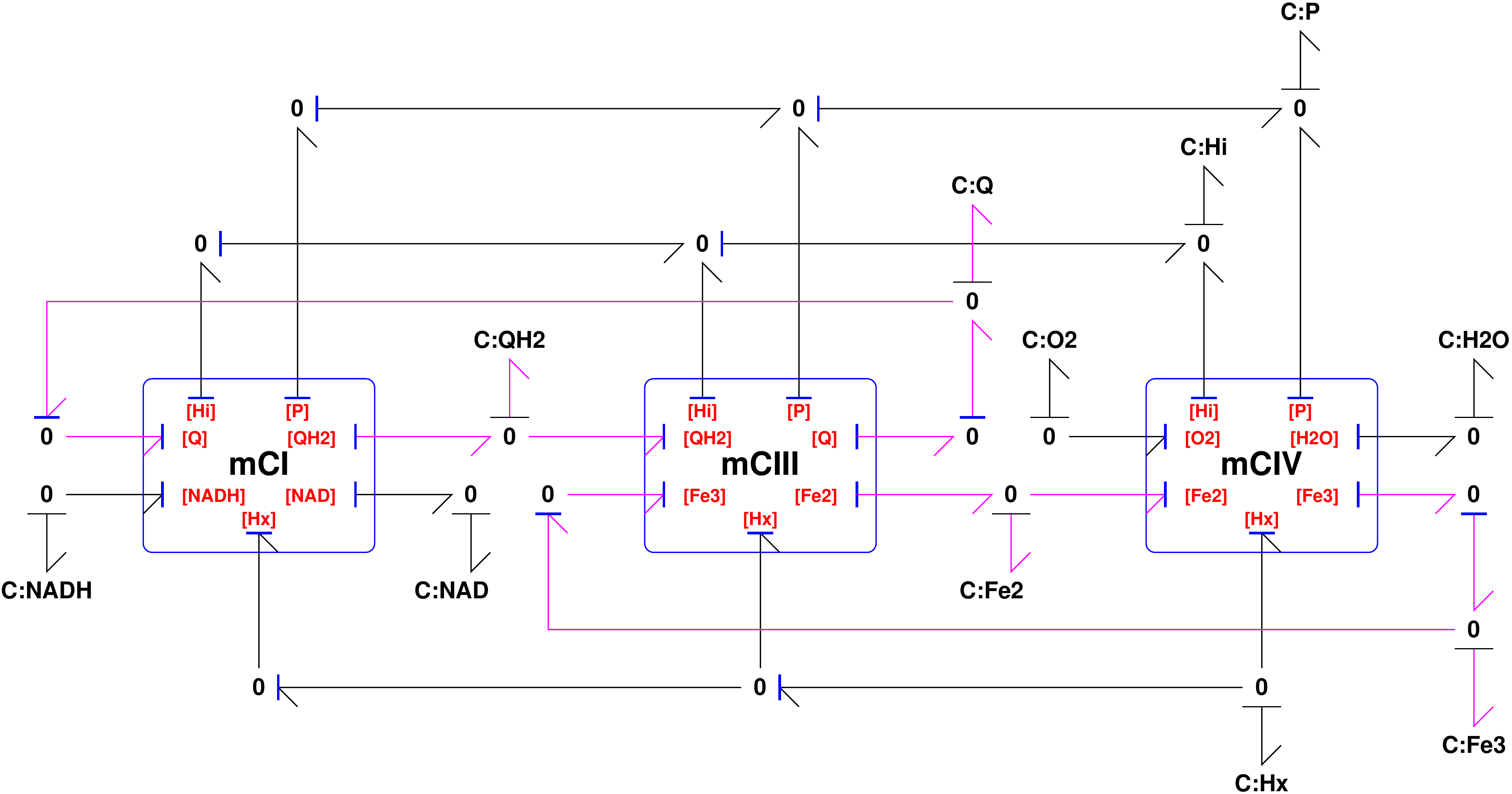}{1}
  \caption[The Electron Transport Chain]{The Electron Transport
    Chain (ETC). 
    The three complexes CI, CIII and CIV are represented by the
    modules mCI (Figure \ref{fig:CI}), mCIII (Figure
    \ref{fig:CIII}) and mCIV (Figure \ref{fig:CIV})
    respectively. 
    All three complexes pump protons $\Hx$ accumulated in \BC{Hx} from
    the matrix across the inner membrane to protons $\Hi$ accumulated
    in \BC{Hi} in the inter-membrane space; the corresponding
    electrical charge is accumulated in \BC{P}. Ubiquone in reduced
    form $\QH$ and oxidised form $\Q$ is recycled around CI \& CIII;
    cytochrome c in reduced form $\Fered$ and oxidised form $\Feox$
    is recycled around CIII \& CIV and the two cycles intersect.}

\label{fig:Ox}
\end{figure}
The mitochondrial electron transport chain (ETC) is a key component of the
bioenergetics of eukaryotic organisms \citep{RicMar10,NicFer13}. The
electrons are transported by, and gain energy from, redox reactions
such as that discussed in \S \ref{sec:redox-reactions} and, as
discussed in \S~\ref{sec:modularity} partially transduce this energy
by pumping protons across the inner membrane to create the
proton-motive force \citep{Mit61,Mit76,Mit11}. 

Because the electron transport chain is primarily concerned with
chemiosmotic  energy transduction, it is natural to use the bond
graph approach for modelling this system. Because the system is
complex, a modular approach enhances understanding. With this in mind,
Figure \ref{fig:Ox} gives a top-level bond graph representation of the
electron transport chain with the following features:
\begin{enumerate}
\item The external substrates are $\NADH$ and $\OO$ and the external
  products are $\NAD$ and $\HHO$ and are represented by \BC{NADH},
  \BC{O2}, \BC{NAD} and \BC{H2O} respectively. As discussed in
  \S~\ref{sec:chemostats}, these components are regarded as
  \emph{chemostats}.
\item The reactions within the three complexes may consume or produce
  protons $\HH$ and thus are dependent on the proton concentration
  often expressed as pH. Protons exist in two separate volumes with
  different pH: the mitochondrial matrix, where they are denoted by
  $\Hx$ and the inter-membrane space where they are denoted by $\Hi$.
\item The three complexes (CI, CIII \& CIV) can be represented as
  redox reactions with an explicit flow of electrons $\ee$ combined,
  as in \S~\ref{sec:modularity}, with a proton pump to transduce the
  electron energy into proton energy by pumping protons across the
  mitochondrial inner membrane
  The matrix protons accumulate in \BC{Hx}, the inter-membrane protons
  accumulate in \BC{Hi} and \BC{P} holds the corresponding electrical
  charge. Because  the mitochondrial inner membrane separating the
  matrix from the intermembrane space has an electrical voltage across
  it, the \BC{P} stores electrical energy. For convenience, the net
  proton charge represented by \BC{P} is denoted $\PP$ within
  reactions to clarify stoichiometry.
\item Ubiquone in reduced form $\QH$ and oxidised form $\Q$ (sometimes
  denoted coenzyme Q or CoQ) is recycled around CI \& CIII; cytochrome
  c in  reduced form%
  \footnote{Following \citep{AtkPau11}, $\Fered$ is used to represent
    reduced cytochrome c (otherwise known as C(red)) and $\Feox$ is
    used to represent oxidised cytochrome c (otherwise known as
    C(ox))}
  $\Fered$ and oxidised form $\Feox$ is recycled around CIII \& CIV and
  the two cycles intersect. As discussed in
  \S~\ref{sec:equalise}, this structure allows the
  Faraday-equivalent potentials to be equalised across the three
  complexes CI, CIII and CIV.
\item Although not included in this paper, the explicit representation
  of electron $\elec$ flow allows electron leakage, and the
  concomitant generation of reactive oxygen species (ROS) such as
  superoxide $\sOO$ and hydrogen peroxide $\HHOO$ (via Superoxide
  Dismutase) \citep{Mur09} to be explicitly modelled.
\end{enumerate}

\subsection{Complex CI}\label{sec:complex-ci}
\begin{figure}[htbp]
  \centering
  \Fig{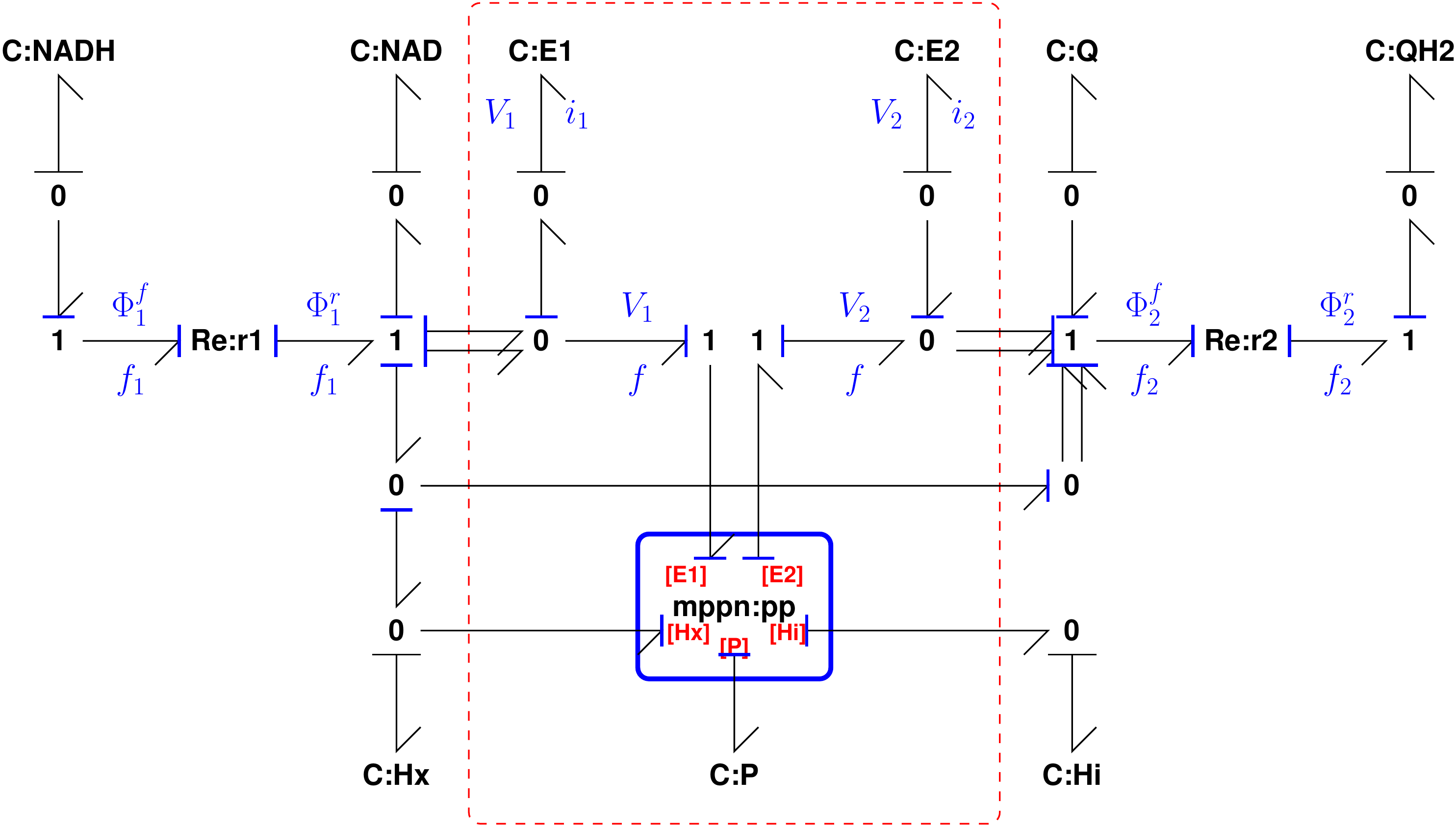}{1}
  \caption{Complex CI}
\label{fig:CI}
\end{figure}
The bond graph of complex CI given in Figure \ref{fig:CI} is
based on the redox reaction of Figure \ref{fig:CIredox} but instead of
the electron-motive force $V_1-V_2$ being dissipated in the resistor
$r$, it is used to drive the proton pump described in
\S~\ref{sec:modularity}.

The proton pump is represented by the bond graph component
\BGL{mppn}{pp} where \BG{mppn} is the modular version of the proton
pump described in Figure \ref{subfig:mppn_cbg} and \texttt{pp}
provides a label indicating a particular instance (in this case with
$n_p=2$).
Following previous notation \citep{Gaw98b,GawBev07,GawCurCra15}, the
five ports are labelled [E1], [E2], [Hx], [Hi] and [P] corresponding to
the five ports designated by the \BSS components of Figure
\ref{subfig:mppn_cbg}.

In a similar fashion to \S~\ref{sec:redox-reactions}, the redox
reaction
\begin{equation}
   \NADH + \Q + \Hx\reac \NAD + \QH
\end{equation}
is split into the two half-reactions:
\begin{align}
  \NADH &\reacu{r1} \NAD + \Hx +2\elec_1\\
  \Q + 2\Hx + 2\elec_2 &\reacu{r2} \QH \label{eq:CI-2}
\end{align}
and these two half reactions are represented by bond graph components
in the same way. There are two differences from the reactions of
\S~\ref{sec:redox-reactions}: 
\begin{enumerate}
\item the energy from the redox reaction is no longer dissipated in
  the component \BRe{r} but are used to drive the proton pump  represented
  by \BGL{mppn}{pp}. The pump removes protons $\Hx$ from the matrix
  and deposits them as $\Hi$ in the intermembrane space; as discussed
  above, the matrix protons accumulate in \BC{Hx}, the inter-membrane protons
  accumulate in \BC{Hi} and \BC{P} holds the corresponding electrical
  charge.
\item the hydrogen ions $\HH_x$ are explicitly associated with the mitochondrial
  matrix.
\end{enumerate}
In this case, $n_p=2$ and thus two protons are pumped from  the
matrix to the intermembrane space for each electron associated with
the redox reaction. As two electrons are associated with each molecule
of $\NADH$, four protons are pumped for each molecule of $\NADH$
consumed in the reaction. In addition, reaction \BRe{r1} produces
one,  and reaction \BRe{r2} consumes two,
protons in the matrix. Thus the overall reaction represented by Figure
\ref{fig:CI} is:
\begin{equation}\label{eq:CI_summary}
   \NADH + \Q + 5\Hx \reacu{CI} \NAD + \QH + 4\Hi + 4\PP
\end{equation}

In Figure 16 of his Nobel Lecture, \citet{Mit93} draws insightful
comparisons between fuel cells and mitochondria. In particular, he notes
that the two half-reactions are coupled by electrons (electricity) and
protons (proticity). The difference is that fuel cells are designed to
generate electricity whereas the electron transport chain of
mitochondria generates proticity.
This is also the situation in Figures~\ref{fig:CI}, \ref{fig:CIII} and
\ref{fig:CIV} where electrons flow in the upper part of the diagram
and protons in the lower and the two half reactions are to the left
and the right. Thus the bond graph representation of complexes CI,
CIII and CIV reflects the situation depicted by \citet[Figure
16]{Mit93}.

\subsection{Complex CIII}
\label{sec:complex-ciii}
\begin{figure}[htbp]
  \centering
  \Fig{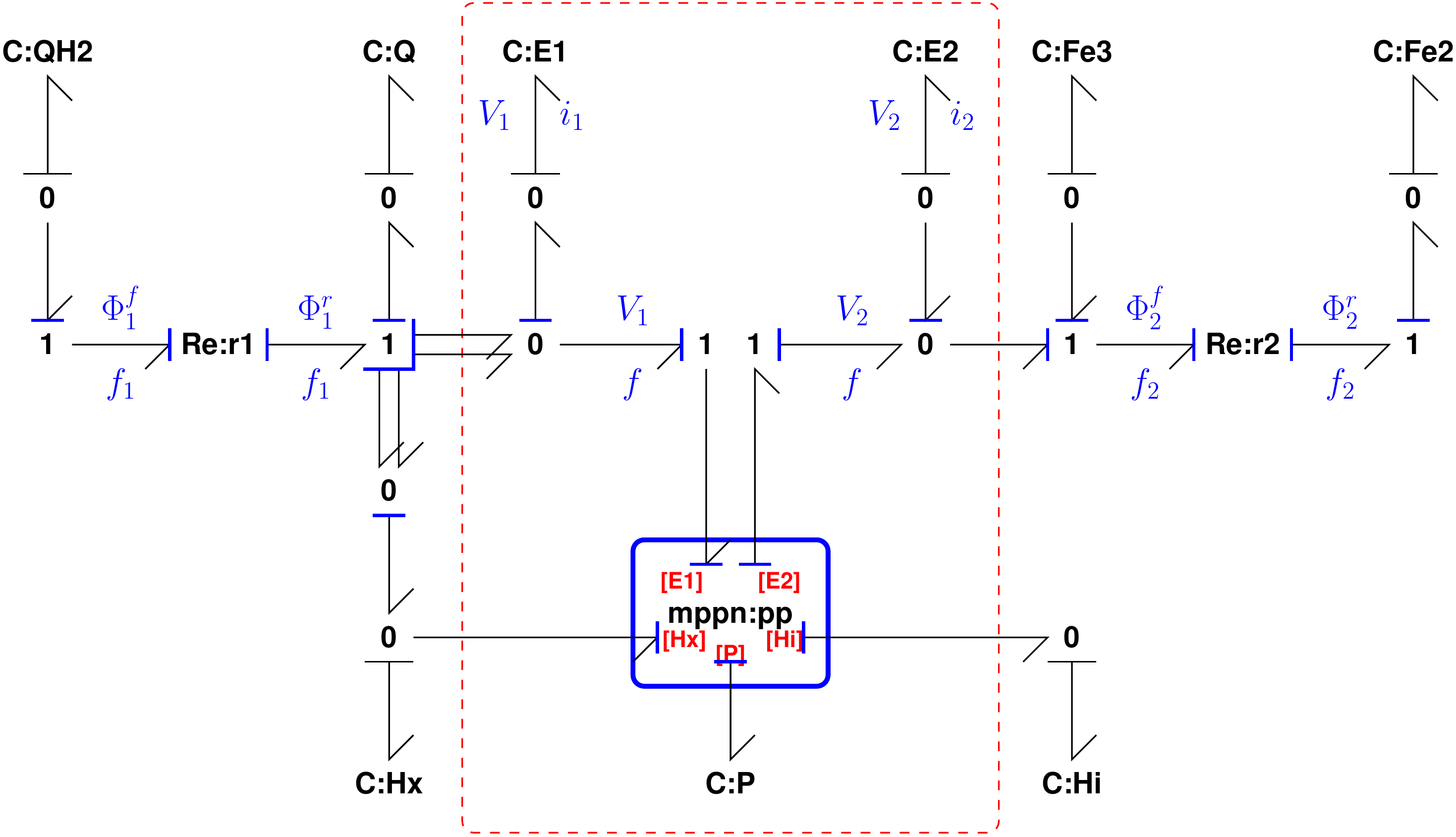}{1}
  \caption{Complex CIII}
\label{fig:CIII}
\end{figure}
The bond graph of complex CIII given in Figure \ref{fig:CIII} is
similar to that of complex CI given in Figure \ref{fig:CI}. The
difference is that it now represents the redox reaction:
\begin{equation}
   \QH + 2\Feox \reac \Q + 2\Fered + 2\Hx
\end{equation}
with the half-reactions
\begin{align}
  \QH &\reacu{r1} \Q + 2\Hx + 2\elec_1\label{eq:CIII-1}\\
   \Feox + \elec_2 &\reacu{r2} \Fered \label{eq:CIII-2}
\end{align}
Note that the first half reaction of CIII \eqref{eq:CIII-1} is the
reverse of the second half reaction of CI \eqref{eq:CI-2}.

As with complex CI, the energy associated with the redox reaction is
used to pump protons across the inner mitochondrial membrane. In this
case $n_p=1$ and thus two protons are pumped from the matrix to the
intermembrane space for each $\QH$ consumed.  
The first half-reaction donates two protons $\Hx$ to the matrix%
\footnote{$\QH$ and $\Q$ exist within the inner membrane. The
  assumption made here is that the the corresponding protons are part
  of the matrix pool. Other assumptions could easily be accommodated
  by modifying Figures \ref{fig:CI} and \ref{fig:CIII}.}
for each $\QH$ consumed.
The first half-reaction produces two electrons for each $\QH$ consumed
and the second half-reaction consumes one electron for each $\Feox$
consumed.  Thus the overall reaction represented by Figure
\ref{fig:CIII} is:
\begin{equation}\label{eq:CIII_summary}
  \QH + 2\Feox \reacu{CIII} \Q + 2\Fered +2\Hi + 2\PP
\end{equation}

\subsection{Complex CIV}
\label{sec:complex-civ}
\begin{figure}[htbp]
  \centering
  \Fig{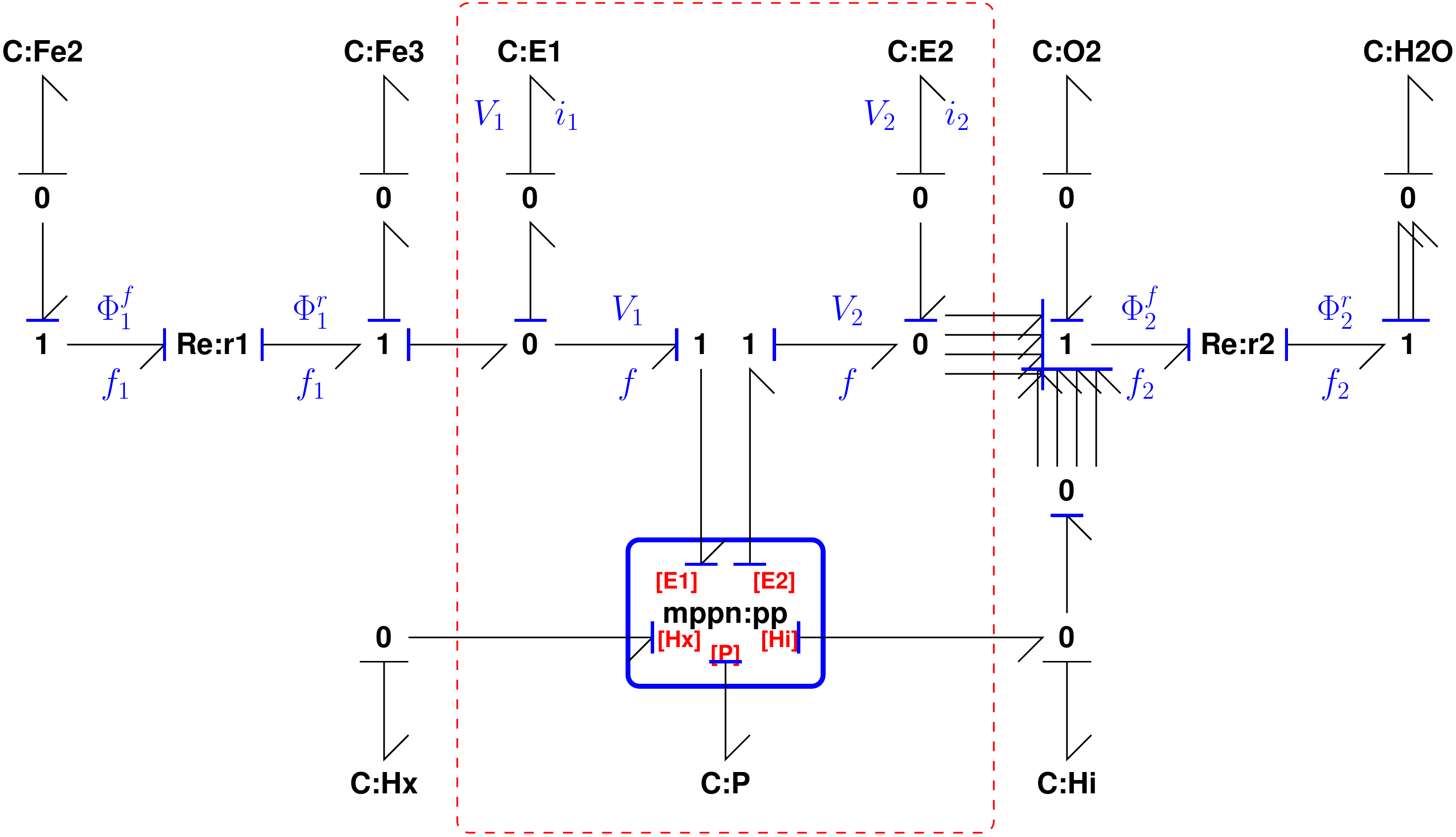}{1}
  \caption{Complex CIV}
\label{fig:CIV}
\end{figure}
The bond graph of complex CIV given in Figure \ref{fig:CIV} is
similar to that of complex CI given in Figure \ref{fig:CI}. The
difference is that it now represents the redox reaction:
\begin{equation}
4\Fered + \OO + 4\Hi \reac 4\Feox + 2\HHO
\end{equation}
with the half-reactions
\begin{align}
\Fered &\reacu{r1} \Feox + \elec_1 \label{eq:CIV-1}\\
 \OO + 4\Hi + 4\elec_2 &\reacu{r2} 2\HHO
\end{align}
Note that the first half reaction of CIV \eqref{eq:CIV-1} is the
reverse of the second half reaction of CIII \eqref{eq:CIII-2}.

As with complex CI, the energy associated with the redox reaction is
used to pump protons across the inner mitochondrial membrane. In this
case $n_p=2$ and thus eight protons are pumped from the matrix to the
intermembrane space for each oxygen molecule $\OO$ consumed.
The first half-reaction consumes two protons $\Hi$ from the
intermembrane space.
Thus the overall reaction represented by Figure
\ref{fig:CIV} is:
\begin{equation}\label{eq:CIV_summary}
  4\Fered + \OO + 8\Hx \reac 4\Feox + 2\HHO + 4\Hi + 8\PP
\end{equation}

Using the methods of \citet{GawCra16X} applied to the bond graphs of
Figures \ref{fig:CI}, \ref{fig:CIII} and \ref{fig:CIV} gives the
following overall chemical equation for the electron transport chain:
\begin{equation}\label{eq:ETC_summary}
  2\NADH+\OO+18\Hx \reac 2\NAD+2\HHO+16\Hi+20\PP
\end{equation}
An alternative approach is to note that
CI \eqref{eq:CI_summary} corresponds to $2\elec$ pumping 4
protons,
CIII \eqref{eq:CIII_summary} corresponds to $2\elec$ pumping 2
protons and
CIV \eqref{eq:CIV_summary} corresponds to $4\elec$ pumping 8
protons. Thus equation \eqref{eq:ETC_summary} arises from multiplying
the stoichiometry of CI and CIII by 2 and adding the resultant equations to
that for CIV; the total number of protons pumped by $4\elec$
passing down the ETC is thus $2{\times}4 + 2{\times}2 + 1{\times}8 = 20$.

Equation \eqref{eq:ETC_summary} is sometimes rewritten with
non-integer stoichiometry as:
\begin{equation}\label{eq:ETC_summary_1}
  \NADH+\frac{1}{2}\OO+9\Hx \reac \NAD+\HHO+8\Hi+10\PP
\end{equation}
The integer stoichiometry version of Equation \eqref{eq:ETC_summary}
is used in the following section.

\section{Energy transduction and affinity equalisation}
\label{sec:equalise}
As an illustration of the potential of the bond graph approach, this
section uses the electron transport chain model of
\S~\ref{sec:electr-transp-chain} to show how the electron-transporting
complexes $\Q/\QH$ and $\Feox/\Fered$ equalise the Faraday-equivalent
potentials along the mitochondrial electron transport chain.
%
\begin{figure}
  \centering
  \SubFig{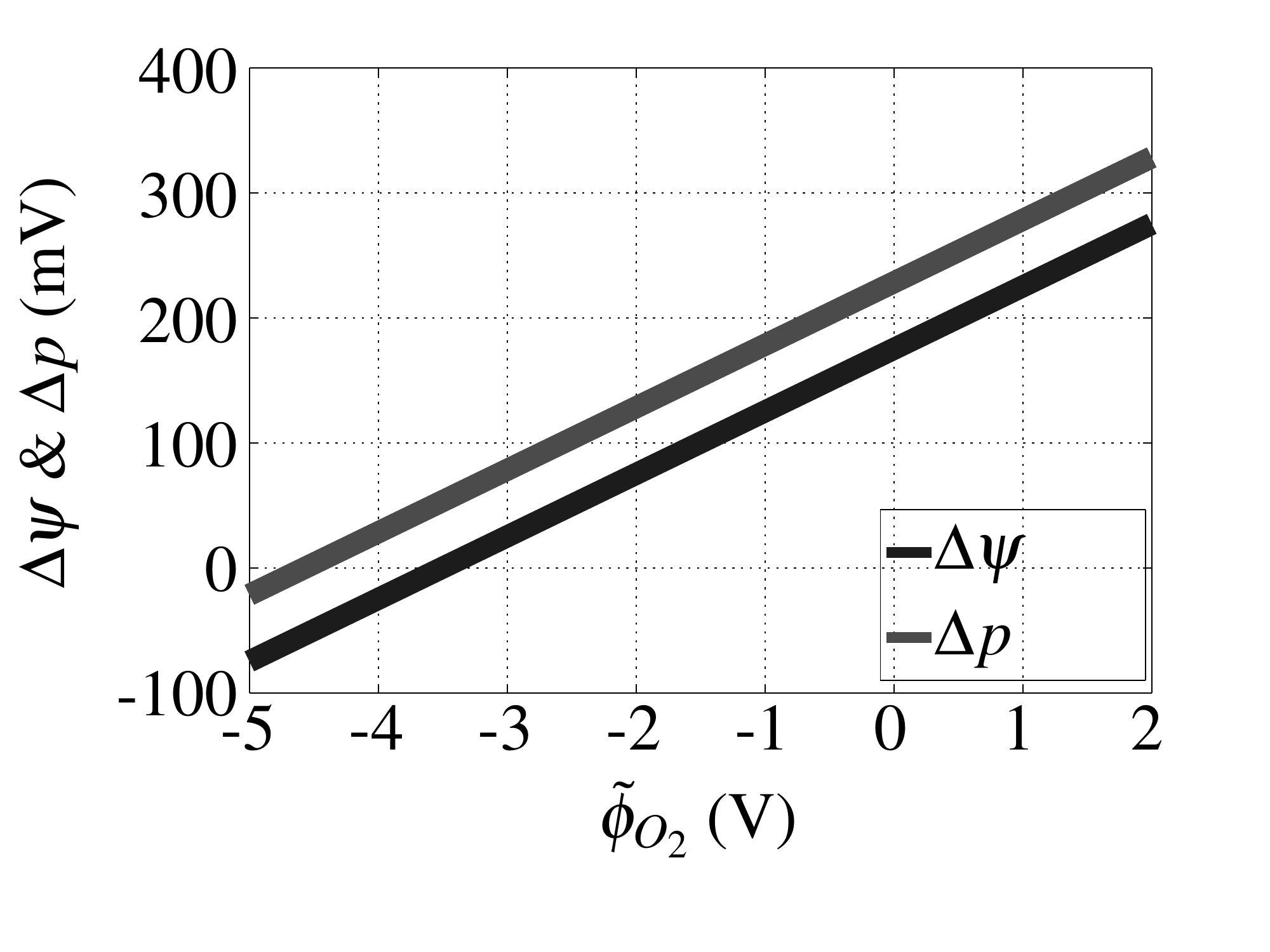}{$\Delta \psi$ \& $\Delta p$}{0.45}
  \SubFig{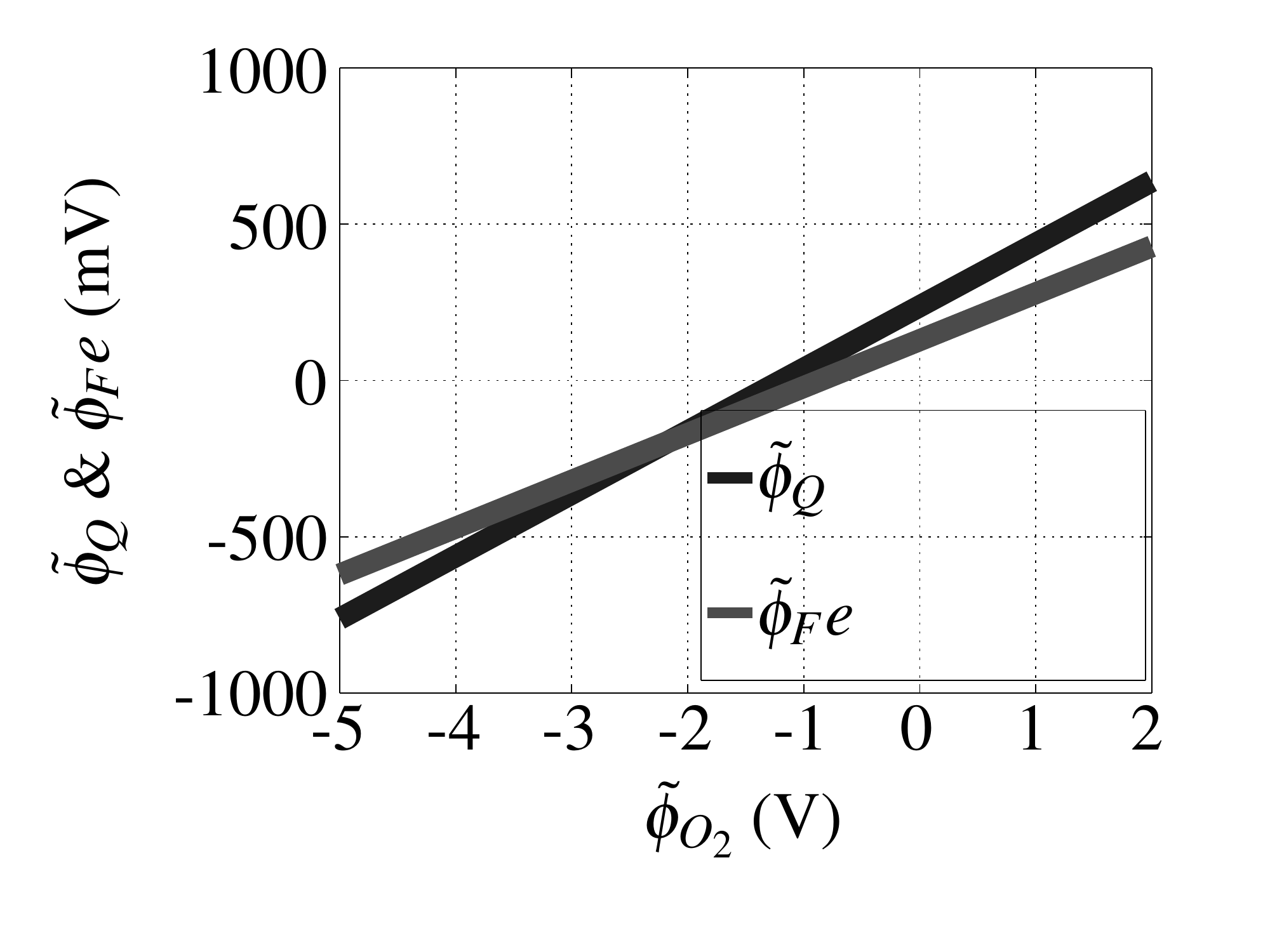}{$\phit_{\Q}$ \& $\phit_{\Fe}$}{0.45}
  \caption[Affinity Equalisation]{Affinity Equalisation}
  \label{fig:Equalisation}
\end{figure}
There are three redox reactions (corresponding to the three
complexes), each of which has two half reactions. However, the first
half-reaction of CIII is a reversed version of the second
half-reaction of CI and the first half-reaction of CIV is a reversed
version of the second half-reaction of CIII. Hence there are just four
half reactions to be considered: those involving $\NADH$, $\Q$, $\Fe$
and $\OO$.

Using Table \ref{tab:FEP}, the four reaction affinities are:
\begin{align}
  \Phi^\std_{\NADH} &= \phi^\std_{\NADH} - \phi^\std_{\NAD}
                      - \phi^\std_{Hx} \notag\\
                    &= 407 - 187  + 460 = \SI{680}{\milli\FV}\\
  \Phi^\std_{Q} &= \phi^\std_{\Q} - \phi^\std_{\QH} + 2 \phi^\std_{\Hx}\notag\\
                    &= 675 + 241 -922 = \SI{-4}{\milli\FV}\\
  \Phi^\std_{\Fe} &= \phi^\std_{\Feox} - \phi^\std_{\Fered}\notag\\
                    &= -67 + 284 = \SI{217}{\milli\FV}\\
  \Phi^\std_{\OO} &= \phi^\std_{\OO}  - 2\phi^\std_{\HHO} +
                    4\phi^\std_{\Hi}\notag\\
                    &= -49 + 4887 - 1628 = \SI{3210}{\milli\FV}
\end{align}
The corresponding complex affinities are
\begin{align}
  \Phi^\std_{CI} &= \Phi^\std_{\NADH} + \Phi^\std_{Q} \notag\\
            &= 680 -4  = \SI{676}{\milli\FV}\label{eq:Phi_CI}\\
  \Phi^\std_{CIII} &= -\Phi^\std_{\Q} + 2\Phi^\std_{\Fe} \notag\\
            &= 4+434 = \SI{438}{\milli\FV}\label{eq:Phi_CIII}\\
  \Phi^\std_{CIV} &= -4\Phi^\std_{\Fe} + \Phi^\std_{\OO}\notag\\
            &= -868+3210 = \SI{2344}{\milli\FV}\label{eq:Phi_CIV}\\
  \Phi^\std_{ETC} &= 2\Phi^\std_{CI}+2\Phi^\std_{CIII}+\Phi^\std_{CIII}\notag\\
            &= 1352 + 876 + 2344 = \SI{4572}{\milli\FV}\label{eq:Phi_Ox}
\end{align}
where $\Phi^\std_{ETC}$ is the overall affinity of the electron
transport chain summarised by Equation \eqref{eq:ETC_summary}%
\footnote{
Equation \eqref{eq:ETC_summary} corresponds to $4\elec$ passing down
the ETC. The non-integer stoichiometry version
\eqref{eq:ETC_summary_1} corresponds to $2\elec$ passing down
the ETC and thus the corresponding affinity is
$\frac{1}{2}\Phi^\std_{ETC} = \SI{2286}{\milli\FV}=\SI{220}{kJ mol^{-1}}$. 
This is the figure quoted in the literature \citep{NicFer13,Nat16}. }
.

Each complex drives protons across the inner membrane against the PMF
$\pmf$; but each complex drives a different number of protons:
$n_{CI}=4$, $n_{CIII}=2$ and $n_{CIV}=8$. The overall number of
protons transferred by the electron transport chain is
\begin{equation}
  n_{ETC} = 2n_{CI}+2n_{CIII}+n_{CIV} = 20
\end{equation}
\begin{align}
  \Phib^\std_{CI} = \frac{\Phi^\std_{CI}}{n_{CI}} &= \frac{676}{4} 
  = \SI{169}{\milli\FV} \label{eq:Phi_CI_prot}\\
  \Phib^\std_{CIII} = \frac{\Phi^\std_{CIII}}{n_{CIII}} &= \frac{438}{2} 
  = \SI{219}{\milli\FV} \label{eq:Phi_CIII_prot}\\
  \Phib^\std_{CIV} = \frac{\Phi^\std_{CIV}}{n_{CIV}} &= \frac{2344}{8} 
  = \SI{293}{\milli\FV} \label{eq:Phi_CIV_prot}\\
  \Phib = \frac{\Phi^\std_{ETC}}{n_{ETC}} &= \frac{4572}{20} 
  = \SI{228}{\milli\FV} \label{eq:Phi_ETC_prot}
\end{align}
Thus with these values of working concentrations, the maximum PMF
$\pmf$ is determined by the smallest of these values, namely
\SI{169}{\milli\FV} for complex CI and the other two complexes have wasted
affinity. 

Complex CIII acts as an electronic bridge  between the the $\Q/\QH$
and the $\Feox/\Fered$ pools and therefore can regulate the entire
electronic transport chain \citep{SarOsy14}. 
In particular, allowing the concentrations relating to the $\Q$ and
$\Fe$ half reactions to vary gives two degrees of freedom to equalise
the complex affinities per proton.  That is:
\begin{align}
  \Phi_{CI} &= \Phi^\std_{CI} + \Phit_{CI} = \Phi^\std_{CI} + \Phit_{\Q}\label{eq:Phi_CI_inc}\\
  \Phi_{CIII} &= \Phi^\std_{CIII} + \Phit_{CIII} = \Phi^\std_{CIII} -
                \Phit_{\Q} + 2\Phit_{\Fe}\label{eq:Phi_CIII_inc}\\
  \Phi_{CIV} &= \Phi^\std_{CIV} + \Phit_{CIV} = \Phi^\std_{CIV} -  4\Phit_{\Fe}\label{eq:Phi_CIV_inc}
\end{align}
As $\Phi_{ETC}$ is not affected by $\Phit_{\Q}$ or $\Phit_{\Fe}$, the
three complex affinities per proton can all be set to
\begin{equation}\label{eq:pmf_ETC}
  \pmf = \bar{\Phi}=\frac{\Phi^\std_{ETC}}{n_{ETC}}
\end{equation}
Hence equations
\eqref{eq:Phi_CI_inc} and \eqref{eq:Phi_CIV_inc} can rewritten as:
\begin{align}
  \Phit_{\Q} &= \Phi_{CI} - \Phi^\std_{CI} = n_{CI}\Phib -
               \Phi^\std_{CI} \notag\\
             &= 912 - 676 =
               \SI{236}{\milli\FV}\label{eq:phit_Q}\\
  \Phit_{\Fe} &= -\frac{1}{4}\lb\Phi_{CIV} - \Phi^\std_{CIV}\rb 
               = -\frac{1}{4}\lb n_{CIV}\Phib - \Phi^\std_{CIV} \rb\notag\\
               &= -\frac{1}{4}\lb1824  - 2344\rb = \SI{130}{\milli\FV}\label{eq:phit_Fe}
\end{align}
It can be verified that this choice of $\Phit_{\Q}$ and $\Phit_{\Fe}$
equalises the three complex affinities per proton:
$\Phib_{CI}=\Phib_{CIII}=\Phib_{CIV}=\Phib$. The corresponding
concentration ratios are:
\begin{align}
  \rho_{\Q} &= \exp \frac{\Phit_{\Q}}{V_N} = 1.0328 \times 10^4\\
  \rho_{\Fe} &= \exp \frac{\Phit_{\Fe}}{V_N} = 151.0
\end{align}

The formulae \eqref{eq:pmf_ETC}, \eqref{eq:phit_Q} \&
\eqref{eq:phit_Fe} can be used to evaluate the PMF $\pmf$, and the
incremental changes in the poll affinities $\Phit_{\Q}$ $\Phit_{\Fe}$
as the concentration of $\OO$, as reflected in the potential
$\phi_{\OO}$ changes.

Figure \ref{fig:Equalisation} shows the results of changing the
concentration of $\OO$ so that the potential is varied by
$\phit_{\OO}$.  Figure \ref{subfig:Ox_pmf} shows how the PMF $\pmf$
and voltage $\emf$ vary and Figure \ref{subfig:Ox_dQFe} shows how the
potentials $\phit_{\Q}$ \& $\phit_{\Fe}$ of the two pools vary. This
corresponds to the fact that hypoxia effects mitochondrial oxidative
metabolism \citep{SolBarLen10}. 

\section{Synthesis of ATP}
\label{sec:ATP}
\begin{figure}[htbp]
  \centering
  \Fig{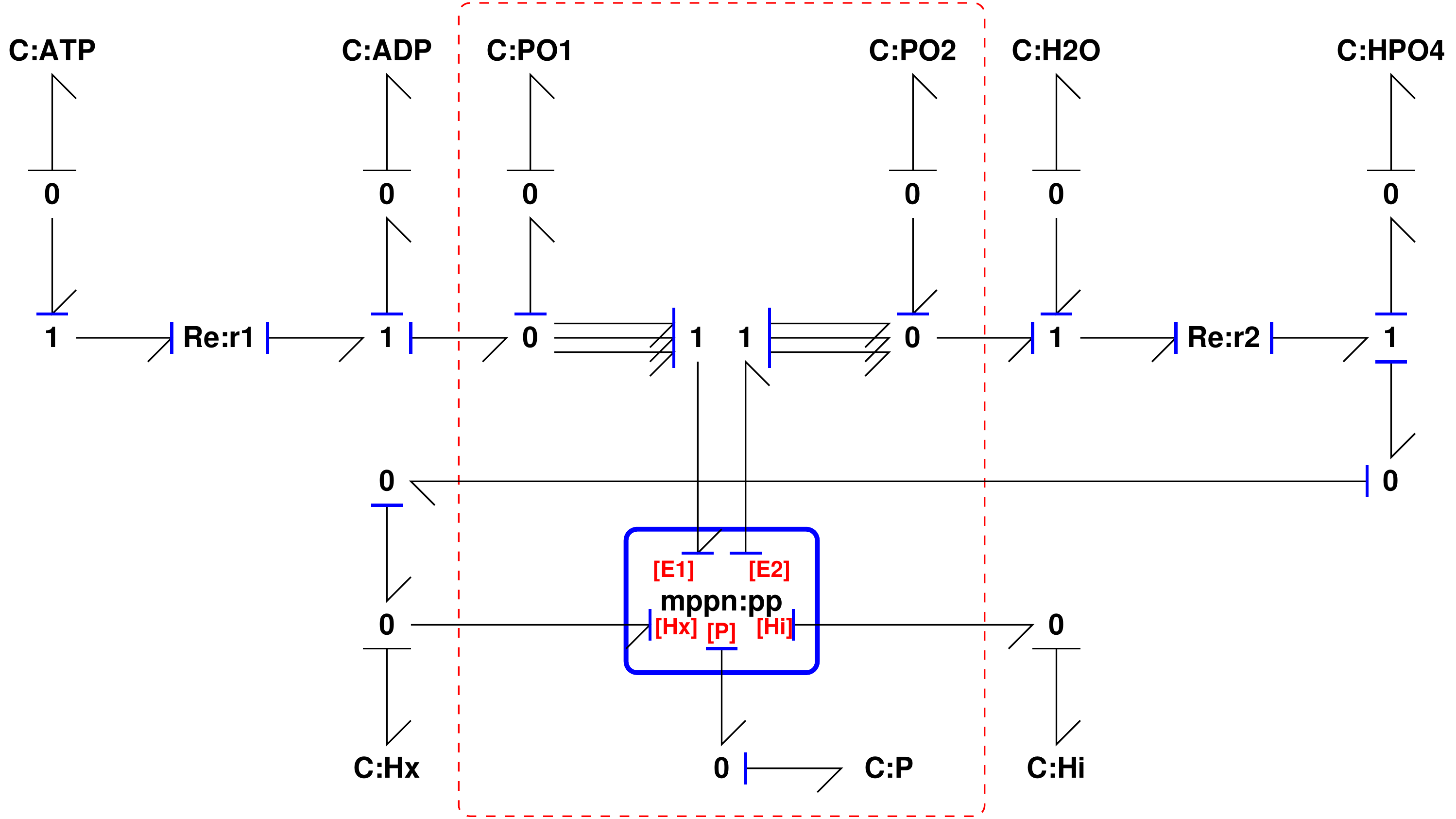}{1}
  \caption{ATP hydrolysis}
  \label{fig:atp-hydrolysis}
\end{figure}
ATP hydrolysis is a key energy-generating reaction in biochemistry.
Following \citet{BerTymStr12}, it may be written as:
\begin{equation}\label{eq:ATP_syn}
  \ATP + \HHO \reac \ADP  + \HPO + \HH
\end{equation}
where $\HPO$ is the orthophosphate ion commonly referred to as
inorganic phosphate or Pi.
Using the proton-motive force generated by the electron transport
chain of \S~\ref{sec:electr-transp-chain}, this reaction can be driven
in reverse to synthesise $\ATP$ from $\ADP$ and $\HPO$.
As discussed in \S~\ref{sec:redox-reactions}, the standard
decomposition of redox reactions into half reactions has a neat bond
graph representation which allows the coupling of a proton pump in a
natural way. The essence of the half-reaction approach is that the two
half reactions are coupled by one or more \emph{electrons}
$\ee$. Here, it is suggested that this idea can be generalised by
allowing the coupling to be some arbitrary chemical entity.  In
particular, although the ATP hydrolysis reaction of Equation
\eqref{eq:ATP_syn} is not a redox reaction, it can be decomposed into
two half reactions as:
\begin{align}
  \ATP  &\reac \ADP + \POOO\label{eq:ATP_syn_1}\\
  \HHO + \POOO &\reac + \HPO + \HH\label{eq:ATP_syn_2}
\end{align}
coupled by the entity $\POOO$.  As discussed in
\S~\ref{sec:complex-ci}, these two half-reactions can be coupled to a
proton pump. In the ATPase complex of vertebrates, three $\ATP$, and
thus three $\POOO$ pump 8 protons \citep[\S~3.6.2]{NicFer13}.
In a similar fashion to the redox reaction representation of
\S~\ref{sec:redox-reactions}, reactions \eqref{eq:ATP_syn_1} and
\eqref{eq:ATP_syn_2} can be represented in bond graph form as in
Figure \ref{fig:atp-hydrolysis}. The overall reaction represented by
Figure \ref{fig:atp-hydrolysis} is:
\begin{equation*}
  3\ATP+3\HHO+5\Hx \reac 3\ADP+3\HPO+8\Hi+8P
\end{equation*}


\begin{figure}[htbp]
  \centering
  \Fig{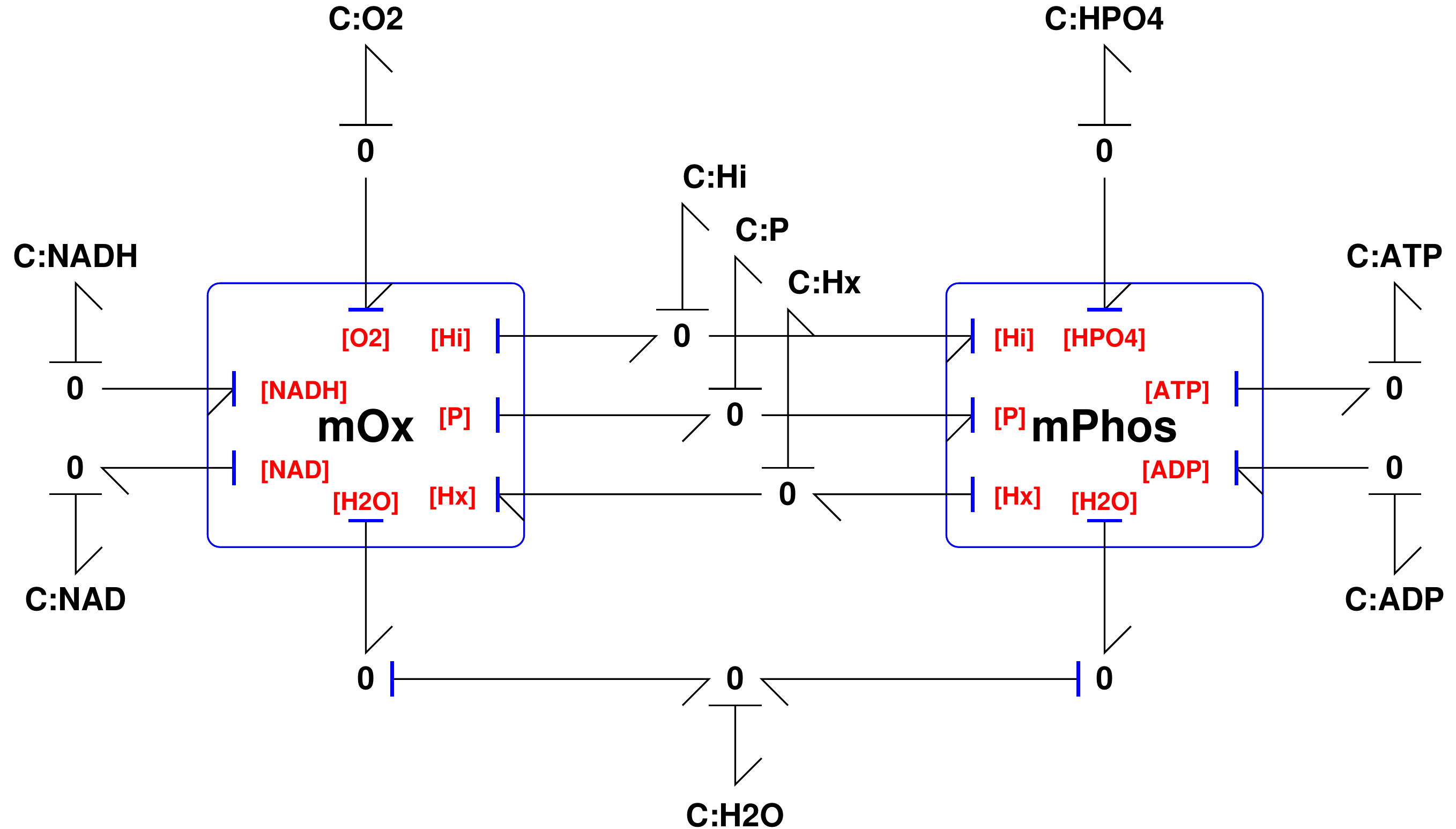}{1}
  \caption{Oxidative phosphorylation. \textbf{mOx} and \textbf{mPhos}
    are the modular versions of the ETC of Figure \ref{fig:Ox} and the
  ATP hydrolysis reaction of Figure \ref{fig:atp-hydrolysis}.}
  \label{fig:OxPhosM}
\end{figure}
Using the approach of \S~\ref{sec:modularity} to modularise the
electron transport chain of Figure \ref{fig:Ox} and the
phosphorylation reaction of Figure \ref{fig:atp-hydrolysis}, oxidative
phosphorylation can be represented in modular bond graph terms as
Figure \ref{fig:OxPhosM}. The coupling of the two modules via the
proton motive force represented by \BC{Hi}, \BC{P} and \BC{Hx} is
clearly visible and the hydrolysis reaction \textbf{mPhos} is driven
in reverse by the ETC \textbf{mOx}.
\section{Conclusion}
\label{sec:conclusion}
It has been shown that combining previous work on the bond graph
modelling of biomolecular systems with the Faraday-equivalent chemical
potential and an alternative concept of bond graph modularity gives a
seamless approach to modelling complex chemiosmotic biological systems
involving biochemical reactions, electrons and protons.
Using a new bond graph representation of redox reactions, the approach
has been applied to give a model of the mitochondrial electron
transport chain. As an illustration, this model is then used to 
show how the electron-transporting complexes $\Q/\QH$ and $\Feox/\Fered$
equalise the Faraday-equivalent potentials along the mitochondrial
electron transport chain.
More generally, the approach of this paper provides an approach
to analysing and understanding energy flows in complex biomolecular
systems -- for example, those within the Physiome Project~\citep{Hun16}.

The appropriate level of complexity of a given model depends on the
use to which the model is put. For example, it would be helpful to
extend the mitochondrial electron transport chain to include the
Q-cycle \citep{HunPalTru03} in complex CIII, the production of
reactive oxygen species (ROS)~\citep{Mur09,VinGri16,BazBeaVin16} and the
corresponding cellular control systems \citep{CosBat12,DunAlvZha15,VinBazBer16}.
On the other hand, for some purposes the model of this paper may be
too detailed; in this case the energy-based pathway analysis of
\citet{GawCra16X} can be used to give a reduced model retaining the
key thermodynamic features.
Versions of a model of a particular biomolecular subsystem (for
example, CIII) can be encapsulated as modules and used and reused
within larger systems to give the appropriate complexity.

It has been argued by \citet{NatVil15} that Mitchell's chemiosmotic
theory is deficient in that ``the energy transducing complexes
involved in oxidative phosphorylation and photosynthesis are
proton-dicarboxylic acid anion cotransporters'' rather than just
proton transporters. It would be interesting to create bond graph
models corresponding to this hypothesis and compare the models with
those of this paper.

The energy balance of biomolecular systems has been discussed in the
literature \citep{GibCha85,HarJolAtt12,SenSte14,GhaNosGol14} and
summarised by \citet{Nat16} in the context of oxidative
phosphorylation.  The energy-based approach used here forms the basis
of an alternative efficiency analysis of biomolecular systems and this
is the subject of current research \citep{GawSeiKam15X}.

Although not discussed here, the bond graph  approach leads to
\emph{dynamic}  models which can be used to generate time-course data
via simulation. Moreover, stability issues can be considered in this
context \citep{GawCra16}. This is the subject of current research.
%
Although not discussed here, spatial variation issues are of
interest. Externally, mitochondria change their shape, size and
clustering configuration \citep{JarGhoDel16} and, according to the
mechano-chemiosmotic model, they change their shape internally
\citep{KasKasKas15}. It would be interesting to include spatial
effects within the bond graph  formulation of this paper.

In addition to the oxidative phosphorylation model of Figure
\ref{fig:OxPhosM}, a model of mitochondrial metabolism would include
glycolysis, the conversion of pyruvate to acetyl coenzyme A, and the
citric acid cycle \citep{BerTymStr12,AlbJohLew15}. The modular energy-based
approach of this paper will be extended to more complete model making
use of the pre-existing modular bond graph model of
glycolysis~\citep{GawCurCra15}.

Because mitochondria are critical to life, mitochondrial dysfunction
is hypothesised to be the source of
ageing~\citep{Wel12,AlbJohLew15}, 
neuro-degenerative diseases \citep{PolCheClo13,Wel12,WelClo12,CloMidWel12,DriSeuSep12,FraGarMid12,MasPrzAbb14}
cancer
\citep{GogOrrZhi08,SolSgaBar11,MarLopGal14} and
other diseases \citep{Wal05,NunSuo12}. 
Although mathematical models of mitochondria exist already
\citep{WuYanVin07,CorAon14,BazBeaVin16,VinBazBer16}, it is hoped that
the engineering-inspired bond graph approach of this paper will shed
further light on the function and dysfunction of mitochondria. This is
the subject of current research.

The equations describing the examples are worked out in some detail in
the paper; however, the results can also be automatically generated
from the system bond graphs.
To illustrate this, a Virtual Reference Environment
\citep{HurBudCra14} is available for this paper at
\doi{10.5281/zenodo.166046}. This contains a ISO image of the
software, bootable by a virtual machine, which not only generates all
figures in the paper but also automatically generates information
about the systems and modules discussed in the paper.

\section{Acknowledgements}
Peter Gawthrop would like to thank the Melbourne School of Engineering
for its support via a Professorial Fellowship, Edmund Crampin for
help, advice and encouragement and Daniel Hurley for help with the
virtual reference environment.
He would also like to thank the reviewers for their suggestions for
improving the paper and drawing his attention to references
\citep{Nat16} and \citep{KasKasKas15}.

\bibliography{common}

\end{document}